\documentclass[usenatbib,usegraphicx]{mn2e}
\usepackage{amsmath}
\usepackage[]{hyperref}
\usepackage{graphicx}
\usepackage[dvipsnames]{xcolor}
\usepackage{units}
\usepackage{ulem}
\usepackage{fixltx2e}
\DeclareGraphicsExtensions{.eps,.pdf,.png,.jpg}

\hypersetup{
    colorlinks=true,
    linkcolor=black,
    citecolor=black,
    filecolor=black,
    urlcolor=black,
}

\begin{document}

\title[Galaxies in Statistical Equilibrium]{On the origin of the fundamental metallicity relation and the scatter in galaxy scaling relations}

\author[J. C. Forbes et al.]{John C. Forbes,$^1$\textsuperscript{\thanks{E-mail: jcforbes@ucsc.edu}}  Mark R. Krumholz,$^1$ Andreas Burkert,$^{2,3}$\textsuperscript{\thanks{Max Planck Fellow}} Avishai Dekel$^4$ \\
$^1$Department of Astronomy \& Astrophysics, University of California, Santa Cruz, CA 95064 USA \\
$^2$University Observatory Munich (USM), Scheinerstrasse 1, 81679 Munich, Germany \\
$^3$Max-Planck-Institut fuer extraterrestrische Physik, Giessenbachstrasse 1, 85758 Garching, Germany \\
$^4$Racah Institute of Physics, The Hebrew University, Jerusalem 91904 Israel
}

\maketitle

\begin{abstract}
We present a simple toy model to understand what sets the scatter in star formation and metallicity of galaxies at fixed mass. The scatter ultimately arises from the intrinsic scatter in the accretion rate, but may be substantially reduced depending on the timescale on which the accretion varies compared to the timescale on which the galaxy loses gas mass. This model naturally produces an anti-correlation between star formation and metallicity at a fixed mass, the basis of the fundamental metallicity relation. We show that observational constraints on the scatter in galaxy scaling relations can be translated into constraints on the galaxy-to-galaxy variation in the mass loading factor, and the timescales and magnitude of stochastic accretion onto star-forming galaxies. We find a remarkably small scatter in the mass loading factor, $\la 0.1$ dex, and that the scatter in accretion rates is smaller than expected from N-body simulations.
\end{abstract}


\section{Introduction}

Large galaxy surveys in the past decade have taught us that star-forming galaxies fall on a main sequence, a tight correlation between their stellar mass and star formation rates. \citep{daddi2007, noeske2007, elbaz2007}. This correlation contains remarkably little scatter - the star formation rate varies by about $\pm0.34$ dex at fixed stellar mass \citep{whitaker2012, guo2013}.

The existence of the star-forming main sequence (MS) and its small scatter have substantially affected our understanding of galaxy evolution. The cosmological paradigm of the past few decades, $\Lambda$CDM, predicts that dark matter halos, and hence the galaxies occupying them, form hierarchically -- large halos are built from mergers of smaller halos. The expectation was therefore that mergers between galaxies would be a major driver of their evolution over cosmological times, frequently triggering starbursts. The small scatter in the main sequence at multiple redshifts has shown, however, that most galaxies are not in fact experiencing any dramatic effects of major mergers \citep{rodighiero2011}, and most stars form in `normal' galaxies lying along this relation.

In addition to the main sequence, galaxies follow other scaling relations. Practically this means that at a given point in cosmic history many properties of galaxies are set by a single parameter associated with the galaxy's mass. This fact has led to the development of a series of models dubbed ``equilibrium'', ``regulator'', or ``bathtub'' models \citep{dekel2009,bouche2010, dave2012, cacciato2012, genel2012, lilly2013, feldmann2013}, in which the properties of a galaxy are self-regulated near a stable equilibrium between accretion of gas, star formation, and outflows. The fundamental ingredients are a star formation rate that increases with increasing gas mass in the galaxy, and a timescale on which the gas reservoir returns to its equilibrium value much shorter than the timescale on which bulk properties of the galaxy (mass, accretion rate, outflow efficiency, star formation timescale, etc.) vary.

Equilibrium models have enjoyed success in heuristically reproducing the average trends in galaxy evolution. However, such models are inherently incapable of quantifying higher-order effects, including scatter in individual scaling relations and various fundamental metallicity relations (FMRs), found by numerous groups \citep{mannucci2010,mannucci2011, lara-lopez2013-b,bothwell2013, stott2013}, typically parameterized as a quadratic function $Z(M_*,$SFR$)$.

Many theoretical studies are focussed on reproducing first-order relations, which is sufficiently difficult in and of itself that second-order relations are often neglected \citep[however, for an exception see][]{dutton2010}. In principle, however, to fully understand galaxy evolution we should be able to understand not only average or median galaxy properties, but their full distribution. A significant advantage of the equilibrium models is that fitting them to any first order relation is trivial, so extending them to understand the higher-order relations is easier than with any other method.

In this work we take one step beyond equilibrium models, allowing the mass accretion rate to vary with a fixed log-normal scatter. Such a scatter is expected based on N-body simulations \citep[e.g.][]{neistein2008} and hydrodynamic simulations \citep[e.g.][]{dekel2013}. Depending on the timescale on which the accretion rate varies, a population of these galaxies may never be in equilibrium. That is, their masses and metallicities may change substantially. However they may still reach a statistical equilibrium, in which the full joint distribution of all galaxy properties becomes time-invariant. 

In section \ref{sec:ms} we introduce the basic formulation of this model and explore the implications for the origin of the width of the star-forming main sequence. We add metallicity to the model in section \ref{sec:Z}, allowing us to examine the FMR and the scatter in the mass-metallicity relation. The model we construct in these sections is independent of the first-order effects considered by most equilibrium models, which allows us to avoid uncertainty in the numerical values of numerous important parameters in galaxy evolution (e.g. the mass loading factor). In section \ref{sec:firstguess} we re-dimensionalize our model, taking a guess at the correct scalings and normalizations to use, and demonstrate the ability of this model to understand unknown physics based solely on the scatter in galaxy scaling relations. We discuss the limitations of our model, quantitative constraints it can place on the accretion process, and alternative explanations for the intrinsic width of these relations in section \ref{sec:discussion}, and conclude in section \ref{sec:conclusion}. 



\section{A very simple model}
\label{sec:ms}
To examine the origin of and scatter within the MS and FMR, we have constructed a minimal model which contains enough physics to produce such features. We do not aim to fully reproduce galaxy properties, tune parameters to match observations, or build upon dark matter merger trees. Instead we aim for simplicity and intuition. In our model we describe the state of a galaxy by two numbers, its cold gas mass $M_g$ and its metalicity $Z=M_Z/M_g$, where $M_Z$ is the mass in metals. The gas mass evolves according to 
\begin{equation}
\label{eq:dMgdt}
\frac{d M_g}{d t} =  \dot{M}_\mathrm{ext}(t) - \frac{M_g}{t_\mathrm{loss}},
\end{equation}
where $t_\mathrm{loss}$ is a the characteristic time over which gas is lost to the system (through the formation of stars and the launching of galactic winds). Here we are implicitly assuming that a given galaxy ejects gas in galactic winds at a rate directly proportional to the star formation rate, the constant of proportionality being defined as the mass loading factor $\eta$. We also assume that all stellar evolution happens instantaneously so that a fixed proportion of all mass which forms stars is immediately returned to the gas reservoir of the galaxy, and a fixed fraction $f_R$ is permanently locked in stellar remnants. We explicitly relate the loss rate with these quantities as follows
\begin{equation}
\label{eq:sfr}
\frac{M_g}{t_\mathrm{loss}} = (f_R+\eta) \dot{M}_\mathrm{SF} =( f_R+\eta) \frac{M_g}{ t_\mathrm{dep}},
\end{equation}
where $\dot{M}_\mathrm{SF}$ is the star formation rate, and $t_\mathrm{dep}$ is the depletion time of all gas in the galaxy. This equation demonstrates that a great deal of (poorly-constrained) physics is hidden in $t_\mathrm{loss}$. In this and the following section, however, we will simply scale the time coordinate to $t_\mathrm{loss}$ and consider ensembles of galaxies with fixed $t_\mathrm{loss}$. This is a powerful technique, since our conclusions in these sections will be independent of the values and scaling relations of $\eta$, $t_\mathrm{dep}$, and so forth.

The external accretion rate is $\dot{M}_\mathrm{ext}$, which we parameterize as a lognormal distribution with fixed median and scatter,
\begin{equation}
\dot{M}_\mathrm{ext}(t) = \exp(\mu + \sigma x(t)),
\end{equation}
where $x(t)$ is a random variable distributed as a standard normal (zero mean, unit variance), with a new value drawn at a fixed time interval $t_\mathrm{coherence}$. A more realistic model might have a spectrum of timescales over which $x$ would vary, but for simplicity, clarity, and analytical tractability we will use a single ``coherence time''.

Equation \ref{eq:dMgdt} is simple enough that it may be solved analytically given a sequence of random numbers, i.e. $x(t)$. It is convenient to first non-dimensionalize by scaling the mass loss rate to the median accretion rate, 
\begin{equation}
\Psi = \frac{M_g}{t_\mathrm{loss}} e^{-\mu},
\end{equation}
and the time to $t_\mathrm{loss}$ via $t=\tau t_\mathrm{loss}$. Thus $\tau$ represents time in units of mass loss timescales, and one may think of $\Psi$ loosely as the star formation rate or the gas mass, though strictly speaking it is the mass loss rate per median accretion rate. The evolution equation then becomes
\begin{equation}
\label{eq:dPsidtau}
\frac{d\Psi}{d\tau} = - \Psi + e^{\sigma x(t)} 
\end{equation}
The value of $\mu$ has entirely dropped out, so the full distribution of $\Psi$ is determined solely by the inherent scatter in the accretion rate, $\sigma$, and the number of mass loss times over which the accretion rate remains constant, $\tau_c \equiv t_\mathrm{coherence}/t_\mathrm{loss}$.

This equation has exactly the same structure as the radiative transfer equation where accretion acts as the source term, time in units of mass loss timescales is similar to the optical depth, and the instantaneous value of $\Psi$ is analogous to the intensity of radiation.

\begin{figure}
\includegraphics[width=9cm]{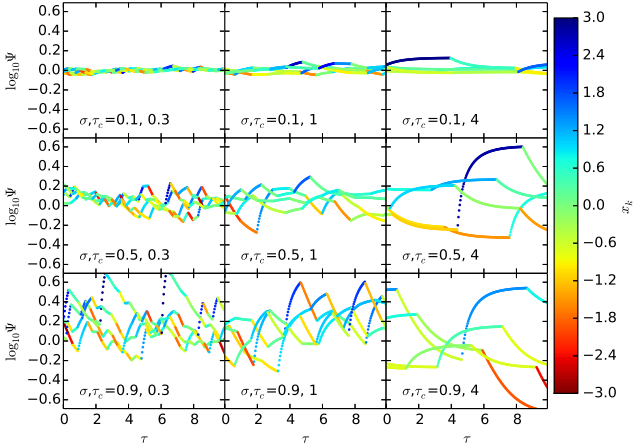}
\caption{$\Psi$ as a function of time. For each pair of $\sigma$ (increasing top to bottom) and $\tau_c$ (increasing left to right), we show the trajectories of five random galaxies over the course of a randomly selected 10 star formation times and colored by the instantaneous value of $x(t)$ -- bluer colors mean higher accretion rates. The galaxies exponentially approach $\Psi = e^{\sigma x_k}$. When $\tau_c$ is short, the many changes in the accretion rate never allow the galaxies to reach the accretion rate -- instead they remain near the average value.}
\label{fig:TimeSeries}
\end{figure}

\begin{figure*}
\includegraphics[width=18cm]{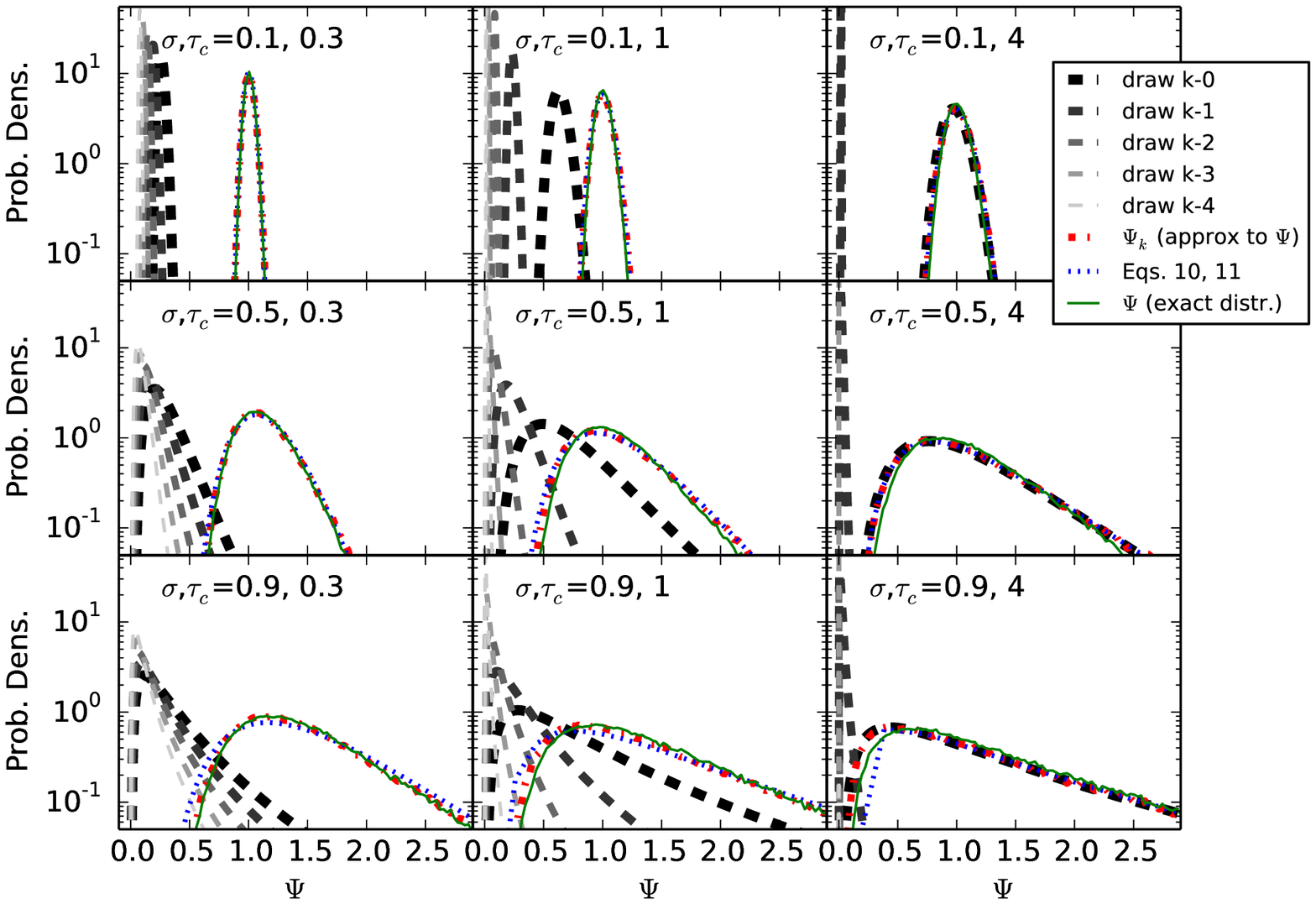}
\caption{The construction of $\Psi$. The red, blue, and green lines show the probability density of a galaxy with a given combination of $\sigma$ and $\tau_c$ having, at a randomly selected time, a given value of $\Psi$. 
 shows the full distribution, red shows the distribution of $\Psi_k$, namely $\Psi$ at a switch in the accretion rate, and blue shows the analytic approximation to $\Psi_k$. These approximations are remarkably good, at least for this selection of $\sigma$ and $\tau_c$. The dashed gray lines show the lognormal distributions from which numbers are drawn and added together to compute $\Psi_k$. As $\tau_c$ increases, fewer draws from the accretion distribution contribute to $\Psi$, until eventually only the last draw matters and all the others are exponentially suppressed (rightmost panel). As the intrinsic width of the accretion distribution increase (top to bottom), the accretion rate distributions increasingly overlap, increasing the chances that the most recent accretion rate does not dominate in determining the current value of $\Psi$.}
\label{fig:ConstructPsi}
\end{figure*}

The mass loss rate as a function of time can now be solved analytically if we are given a sequence of standard normals, $x_k$, $k=0,...$, where each $k$ corresponds to a new draw from the lognormal accretion rate distribution. In particular, suppose we know $\Psi_k$, the value of $\Psi$ at the time that a new accretion rate is drawn. The value of $\Psi$ from then up to the subsequent draw is given by
\begin{equation}
\label{eq:mgrel}
\Psi(\Delta) = \Psi_k e^{-\Delta} + e^{\sigma x_k}(1-e^{-\Delta})\ \mathrm{for }\ 0\le\Delta\le\tau_c
\end{equation}
where we have defined $\Delta \equiv \tau - k \tau_c$, the time since the most recent switch in the accretion rate. This is simply a relaxation equation. We can see this more explicitly by defining $r_k = \Psi_k/e^{\sigma x_k}$, the mass loss rate at the time of the switch in accretion rate relative to the newly-chosen accretion rate -- this is a measure of how far from equilibrium the system is immediately after a switch. Large values mean that  the galaxy is losing much more mass than it is accreting, values less than unity mean gas is building up, and $r_k=1$ means that the galaxy has equilibrated to its current accretion rate. Note that $r_k>0$. Using $r_k$, we see that
\begin{equation}
\Psi(\Delta) / e^{\sigma x_k} = 1 + e^{-\Delta} (r_k-1) \ \mathrm{for }\ 0\le\Delta\le\tau_c
\end{equation}
In other words, the loss rate approaches the accretion rate exponentially on a mass loss timescale, with the deviation determined by the deviation at the time the accretion rate switched ($\Delta=0$).

Using equation \ref{eq:mgrel}, we can recursively compute $\Psi_{k+1}$ from the previous $\Psi_k$ by setting $\Delta=\tau_c$,
\begin{equation}
\label{eq:recursion}
\Psi_{k+1} = \Psi_k e^{-\tau_c} + e^{\sigma x_k} (1-e^{-\tau_c})
\end{equation}
The sequence of $x_k$'s thereby determine a sequence of $\Psi_k$'s, which are used as a scaffolding to construct the full solution $\Psi(\tau)$ for a given realization of the random variables. Examples of such realizations for various values of $\sigma$ and $\tau_c$ are shown in figure \ref{fig:TimeSeries}.

We may also solve the recursion relation (equation \ref{eq:recursion}) for $\Psi_k$ explicitly, again given the sequence $x_k$.
\begin{equation}
\label{eq:psik}
\Psi_k = (1 - e^{-\tau_c}) \sum_{i=0}^{k-1} e^{-(k-i-1) \tau_c} e^{ \sigma x_i} + \Psi_0 e^{-k\tau_c} 
\end{equation}
We see that the mass loss rate at a switch in the accretion rate is therefore simply a sum of lognormally-distributed random variables. Each draw from the distribution loses influence as it recedes into the past. In particular $(k-i)\tau_c$ is simply the number of loss times since $x_i$ was drawn. The prefactor of $1-e^{-\tau_c}$ accounts for the fact that each individual draw of the accretion history matters less as the coherence time gets shorter. For example, even if the accretion rate is very large, if the coherence time is very short, the galaxy will only experience that accretion rate for a very short time.

In the limit that $\tau_c \gg 1$, only the most recent draw from the distribution matters and all previous draws are exponentially suppressed. In the opposite limit, $\tau_c \ll 1$, the leading factor $1-e^{-\tau_c} \rightarrow \tau_c \approx N_\mathrm{loss}^{-1}$, where $N_\mathrm{loss}$ is the number of draws from the accretion rate distribution in a given mass loss time. For very short coherence times, $\Psi_k$ becomes an average of the lognormal accretion rates over the most recent mass loss timescale, with more recent accretion rates weighted somewhat more heavily. The full probability distribution of $\Psi$ and $\Psi_k$ (which serves as an approximation to $\Psi$) for various values of $\tau_c$ and $\sigma$ are shown in figure \ref{fig:ConstructPsi}, along with the probability distributions of the 5 most recent draws from the appropriate lognormal distributions which are added together to give $\Psi_k$. As the coherence times grow longer (panels from left to right), fewer draws from the accretion rate contribute to the current mass loss rate. As the intrinsic scatter in the accretion rate increases (top to bottom), the probability density of the draws from the accretion rate overlap more, meaning that it is increasingly likely that the most recent accretion rate is not the largest contributor to the current mass loss rate.

\begin{figure}
\includegraphics[width=9cm]{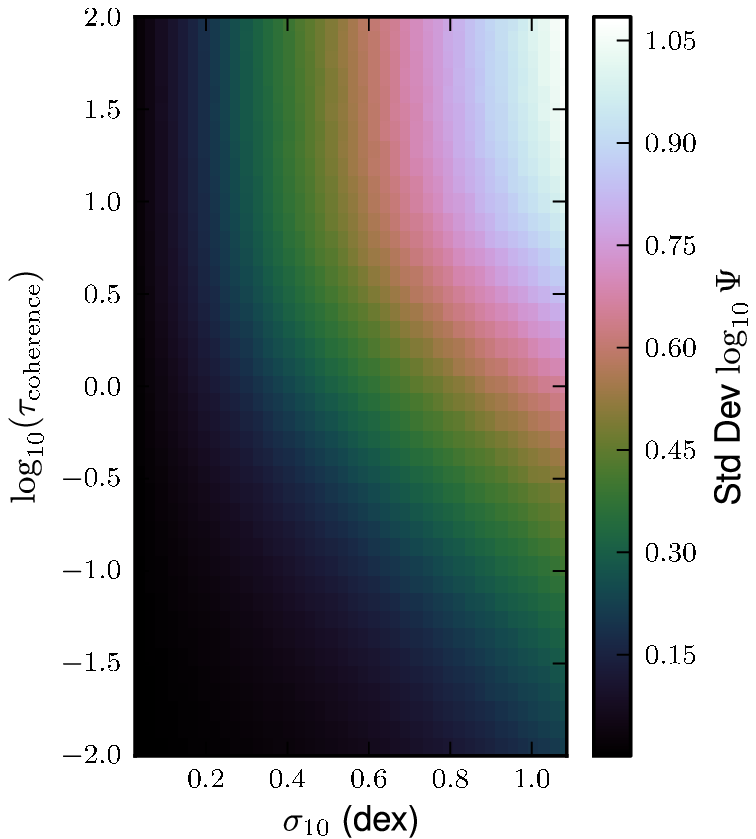}
\caption{The width of the ``main sequence''. Each pixel represents an ensemble of galaxies with fixed $\tau_c$ and $\sigma$, wherein the mass loss rate $\Psi$ was measured for each galaxy at a random time. The standard deviation of $\log\Psi$ in each bin is plotted above. Longer coherence times (upwards in the plot) allow the galaxies to equilibrate so that $\Psi$ approaches the accretion rate, so the scatter in $\Psi$ approaches $\sigma$. Coherence times shorter than the mass loss timescale, i.e. $\log_{10} \tau_c \la 1$, lead to a reduction in the scatter roughly in proportion to $\tau_c$ -- individual draws from the accretion rate matter increasingly less, and the galaxies do not have enough time for $\Psi$ to approach the accretion rate before it changes.}
\label{fig:sigSF}
\end{figure}

The distribution of a random variable, i.e. $\Psi_k$ which is the sum of lognormally distributed random variables, i.e. $(1-e^{-\tau_c})e^{-j\tau_c + \sigma x_j}$, may itself be approximated as a lognormal random variable \citep{fentonlawrence1960} with variance (in the log)
\begin{equation}
\label{eq:width}
\sigma_{\Psi_k}^2 \approx \ln\left[1+ \left(e^{\sigma^2} -1 \right) \frac{1 - e^{-\tau_c}}{1 + e^{-\tau_c}}  \right].
\end{equation} 
and median
\begin{equation}
\mu_{\Psi_k} \approx \sigma^2/2  - \sigma_{\Psi_k}^2/2.
\end{equation}
Equation \ref{eq:width} represents a good analytical guess for the width of the star-forming main sequence. In the limit of large $\tau_c$, $\sigma_{\Phi_k} \rightarrow \sigma$ as expected -- when the galaxies are able to equilibrate to their accretion rates, the width of the main sequence is simply the intrinsic width in the accretion rate, $\sigma$. In this limit $\mu_{\Phi_k} \rightarrow 0$, i.e. the median mass loss rate approaches the median accretion rate.

In the limit of short $\tau_c$, the fraction $(1-e^{-\tau_c})/(1+e^{-\tau_c})$ reduces to $\tau_c/2$. Thus when $\sigma$ is also sufficiently small (namely $\sigma^2 \ll 1$), the variance of $\Psi_k$ reduces to $\sigma^2 \tau_c/2$. Physically, when the accretion rate varies rapidly compared to the mass loss time, any individual draw from the accretion rate distribution becomes unimportant and all that matters is the long-term average. The exception is when $\sigma$ is large, in which case extremely large accretion rates become common and $\sigma_{\Psi_k}$ again approaches $\sigma$, though we caution that the distribution becomes less log-normal as $\sigma$ increases with small $\tau_c$. Figure \ref{fig:sigSF} shows $\sigma_{\Psi_k}$ computed with a Monte Carlo simulation (for details see appendix \ref{app:mc}).

From this discussion, we see that $\tau_c$ and $\sigma$ compete in setting the width of the main sequence, even after $\sigma_{\Psi_k}$ has been scaled by the intrinsic accretion rate. This suggests that $\sigma$ may be interpreted as a third timescale in the problem, namely the number of star formation times necessary to forget a typical accretion event.

\section{Including metallicity}
\label{sec:Z}
The metal content of the galaxy is evolved according to the instantaneous recycling approximation, in the spirit of \citet{tinsley1980} and \citet{maeder1992},
\begin{equation}
\frac{d M_g Z}{d t} = Z_{IGM}\dot{M}_\mathrm{ext} + \dot{M}_\mathrm{SF}(f_R y - f_R M_Z/M_g - \eta Z_w)
\end{equation}
New metals are added along with accreting matter (the first term), and respectively produced by, locked up in the products of, and ejected from the galaxy by, star formation. 

The star formation rate, taking into account galactic winds and stellar evolution (see equation \ref{eq:sfr}), is $\dot{M}_\mathrm{SF} =(f_R+\eta)^{-1} M_g/t_\mathrm{loss}$. The yield $y$ is defined as the mass of metals produced during the course of stellar evolution per unit mass of gas locked in stellar remnants -- if 1000 $M_\odot$ of gas forms stars, a total of $y f_R 1000 M_\odot$ of metals will be produced by these stars and returned to the ISM. Following \citet{forbes2013}, we parameterize the wind metallicity $Z_w$ as 
\begin{equation}
Z_w = Z + \xi y f_R / \max(\eta, 1-f_R).
\end{equation}
The usual assumption throughout the literature on chemical evolution is that $Z_w = Z$, i.e., before gas is ejected from a galaxy by stellar feedback, it is assumed to be perfectly well-mixed with the ambient ISM \citep[for recent exceptions see][]{peeples2011,krumholz2012,vogelsberger2013}. This is a strong assumption because galactic winds are likely to be preferentially metal-enriched. Physically this is because metals are produced in the same places, sometimes by the same events, which are likely to cause galactic-scale outflows, namely sites of recent star formation, where massive stars emit ionizing radiation and end their lives as supernovae. The assumption that $Z_w=Z$ corresponds in our model to $\xi=0$ -- the other extreme value is $\xi=1$, corresponding to exactly no mixing between the metal-rich ejecta of stars and the ISM.

Non-dimensionalizing as in the previous section, we arrive at  
\begin{equation}
\label{eq:dMZdt}
\frac{d \Psi Z}{d \tau} = - \Psi Z  + Z_{IGM} e^{\sigma x(t)} + q \Psi
\end{equation}
where all of the factors associated with star formation and feedback can be collected into a single parameter
\begin{equation}
q \equiv \frac{y f_R}{f_R + \eta} \left(1 - \frac{\eta \xi}{\max(1-f_R,\eta)}\right).
\end{equation}
Much of the uncertainty in modeling the metallicity evolution of galaxies is encapsulated in $q$. We note that in the limit that $\eta$ is large (probably the case for low-mass galaxies), $q\rightarrow y f_R (1-\xi)/\eta$, while for small values of $\eta$, $q\rightarrow y$. In this sense $q$ may be considered an effective yield, and the more important galactic winds are, the more uncertain this parameter becomes.

Combining equations \ref{eq:dMZdt} and \ref{eq:dMgdt}, we can obtain the evolution equation for metallicity,
\begin{equation}
\frac{d Z}{d\tau} = \frac{Z_{IGM} - Z}{\Psi/e^{\sigma x}} + q
\end{equation}
This equation represents a competition between the effective yield of new metals formed during the course of stellar evolution and dilution of metals by accretion. In equilibrium, namely when the accretion rate is constant for many mass loss times, $\Psi/e^{\sigma x} \rightarrow 1$, so for $dZ/d\tau = 0$, we find that the equilibrium metallicity is
\begin{equation}
Z_{eq} = q + Z_{IGM}. 
\end{equation}
The corresponding quantity for the relative mass loss rate is $\Psi_{eq} = e^{\sigma x}$, where $x$ is a standard normal. In this sense, metallicity is very different from $\Psi$ -- regardless of the accretion rate, the metallicity approaches a constant value, whereas the mass loss rate approaches whatever random value of the accretion rate it is being fed.

\begin{figure}
\includegraphics[width=9cm]{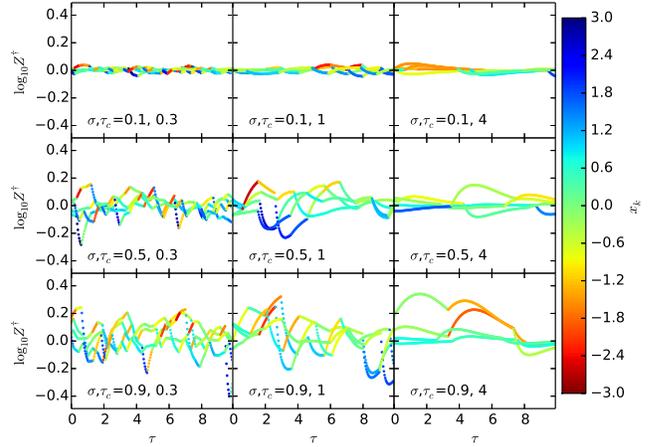}
\caption{$Z^\dagger$ as a function of time. For each pair of $\sigma$ (increasing top to bottom) and $\tau_c$ (increasing left to right), we show the trajectories of five random galaxies over the course of a randomly selected 10 star formation times and colored by the instantaneous value of $x(t)$ -- bluer colors mean higher accretion rates. We can immediately see that high accretion rates tend to lead to low metallicities. We also see that for large coherence times (rightward) each galaxy reaches an extremum in metallicity before returning to its equilibrium value ($Z^\dagger = 1)$}
\label{fig:TimeSeriesZ}
\end{figure}

The factor of $q$ can be factored out of the metallicity evolution equation by defining
\begin{equation}
Z^\dagger = \frac{Z -Z_{IGM}}{q},
\end{equation}
the metallicity scaled to its equilibrium value by the effective yield -- $Z_{eq}$ corresponds to $Z^\dagger = 1$. The evolution equation for $Z^\dagger$ then becomes independent of $q$, yielding
\begin{equation}
\label{eq:dZddt}
\frac{d Z^\dagger}{d\tau} = \frac{-Z^\dagger }{1+e^{-\tau} (r_k-1)} + 1
\end{equation}
The denominator is recognizable as the mass loss rate normalized by the current accretion rate, $e^{\sigma x_k}$. 

As with the star formation equation, we may solve equation \ref{eq:dZddt} by elementary methods for a given fixed value of $r_k$, i.e. during the period when the accretion rate is constant. The solution is
\begin{equation}
\label{eq:Zdt}
Z^\dagger(\Delta) = \frac{Z^\dagger(0) r_k + e^\Delta-1 +(r_k-1)\Delta}{e^\Delta-1+r_k} \ \ \mathrm{for}\ 0\le\Delta\le\tau_c
\end{equation}
In the limit of large $\Delta$, we recover that $Z^\dagger \rightarrow 1$, i.e. the metallicity returns to its equilibrium value, though depending on how different the initial mass loss rate is from the accretion rate, $\Psi_k/e^{\sigma x_k} = r_k$,  the value of $Z^\dagger$ may change dramatically over the course of a star formation time ($\Delta \sim 1$) before exponentially returning to unity. Examples of such trajectories are shown in figure \ref{fig:TimeSeriesZ}. Notice that in the right-most panels, despite the somewhat large value of $\tau_c$, there are significant deviations from equilibrium immediately following a change in the accretion rate. For example, if the accretion rate increases, the ISM begins getting diluted immediately by the large accretion rate. The star formation rate, and hence the rate of metal production, takes some time to adjust to the new accretion rate while gas builds up in the ISM. It therefore takes some time before metal production can `pollute' the large supply of low-metallicity infalling gas and return to the equilibrium $Z^\dagger=1$, wherein for every unit of new clean gas added to the galaxy, star formation provides enough mass in metals to bring it to $Z=q+Z_{IGM}$. Similarly, if the accretion rate decreases suddenly, the rate at which metals are produced exceeds the rate necessary to pollute the new, smaller, supply of accreting gas, and the metallicity of the ISM increases until star formation and galactic winds can burn through the gas reservoir, allowing the galaxy to adjust to a lower rate of star formation (and hence metal production).

Immediately after a change in the accretion rate, i.e. in the limit that $\Delta \ll 1$, 
\begin{equation}
Z^\dagger(\Delta) \approx Z^\dagger(0) + (1 - Z^\dagger(0)e^{\sigma x_k} / \Psi_k) \Delta 
\end{equation}
Thus we can see that immediately following a change in accretion rate, whether the metallicity will increase or decrease depends only on the ratio $Z^\dagger(0) e^{\sigma x_k}/\Psi_k$. When this number is greater than unity, because the metallicity is high and/or the new accretion rate is larger than the present mass-loss rate, the metals will be diluted, whereas if this number is less than one, e.g. in a low-metallicity galaxy and/or one facing a sudden drop in accretion rate, stars will be forming fast enough to pollute the gas reservoir. This relation demonstrates the basic physical mechanism which gives rise to the the mass-star formation-metallicity plane in our model, but also shows that this is a statistical relation only. A higher accretion rate (and hence an increasing star formation rate) is not guaranteed to produce a lower metallicity

We can solve for the maximum deviation from $Z^\dagger=1$ if the galaxy begins in equilibrium, namely $Z^\dagger(0)=1$. In this case,
\begin{equation}
Z^\dagger_\mathrm{extreme} = \frac{(r_k -1)(1 + W_0((r_k-1)/e)}{r_k -1 + \exp(1 + W_0((r_k-1)/e))}
\end{equation}
where $W_0(x)$ is the Lambert W function, namely the real solutions to $y=W_0(y)e^{W_0(y)}$.

\begin{figure*}
\includegraphics[width=18cm]{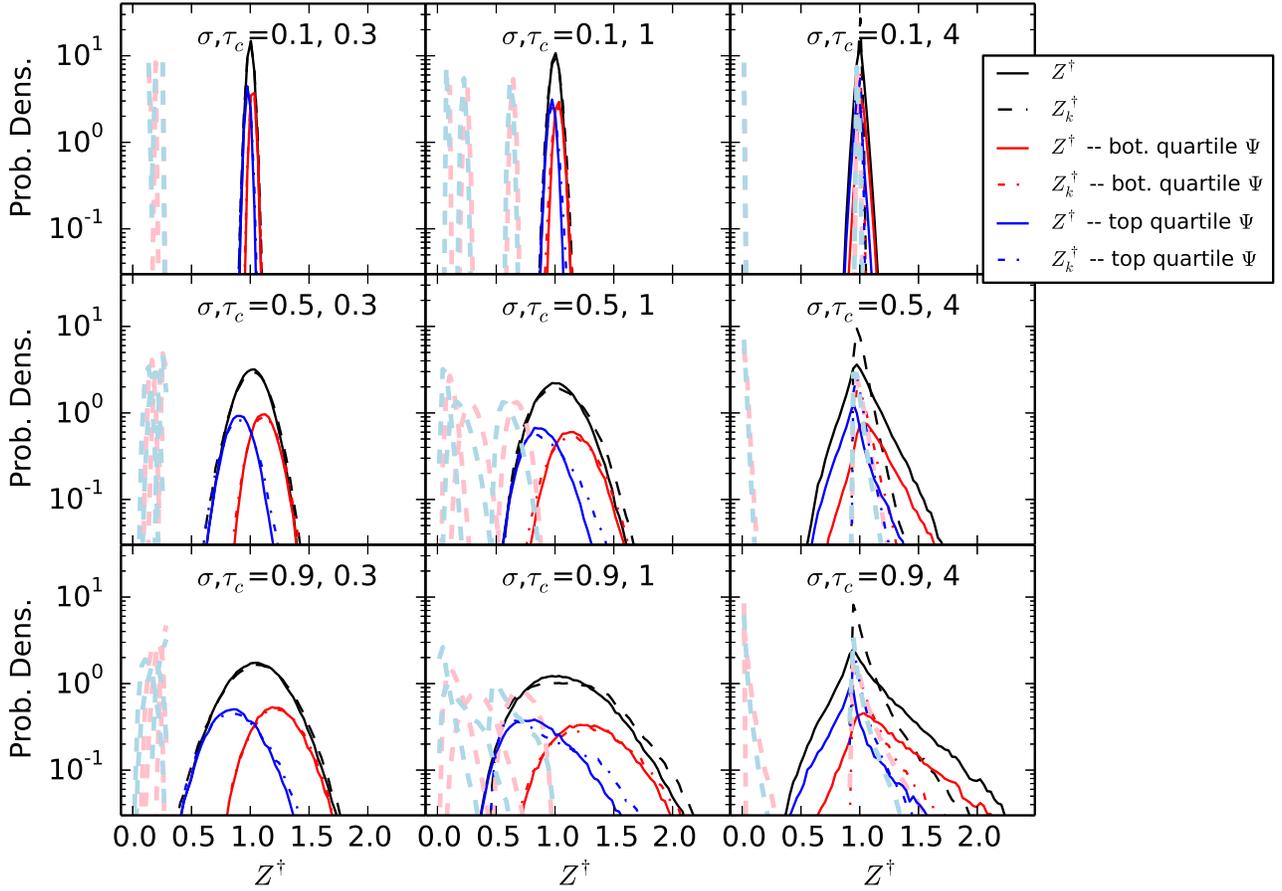}
\caption{The construction of $Z^\dagger$. For the same parameters shown in the previous figures, we plot the PDF of $Z^\dagger$ (solid line), the approximation to it using $Z^\dagger_k$ (equation \ref{eq:Zdk}), and we also split the probability distribution into galaxies with high (low) star formation rates in blue (red).  We can immediately see that high metallicity galaxies tend to have low star formation rates, though the probabilities do overlap substantially. Light blue and red lines show the PDF of the contributions to $Z^\dagger_k$, split into the top and bottom quartiles in star formation rate, from old terms in the summation of equation \ref{eq:Zdk}. As with $\Psi$, the probability distribution of $Z^\dagger$ may be thought of as the sum of a series of random draws wherein the influence of older draws is exponentially forgotten, although galaxies which end up in the blue (red) bin at the present time are likely to have had preferentially higher (lower) SFRs in the past, at least for $\tau_c \la 1$. Unlike with $\Psi$, $Z^\dagger$ becomes noticeably non-log-normal for larger values of $\tau_c$, since galaxies sampled at a random time are likely to be near the equilibrium value $Z^\dagger=1$, regardless of the accretion rate.}
\label{fig:ConstructZ}
\end{figure*}

\begin{figure}
\includegraphics[width=9cm]{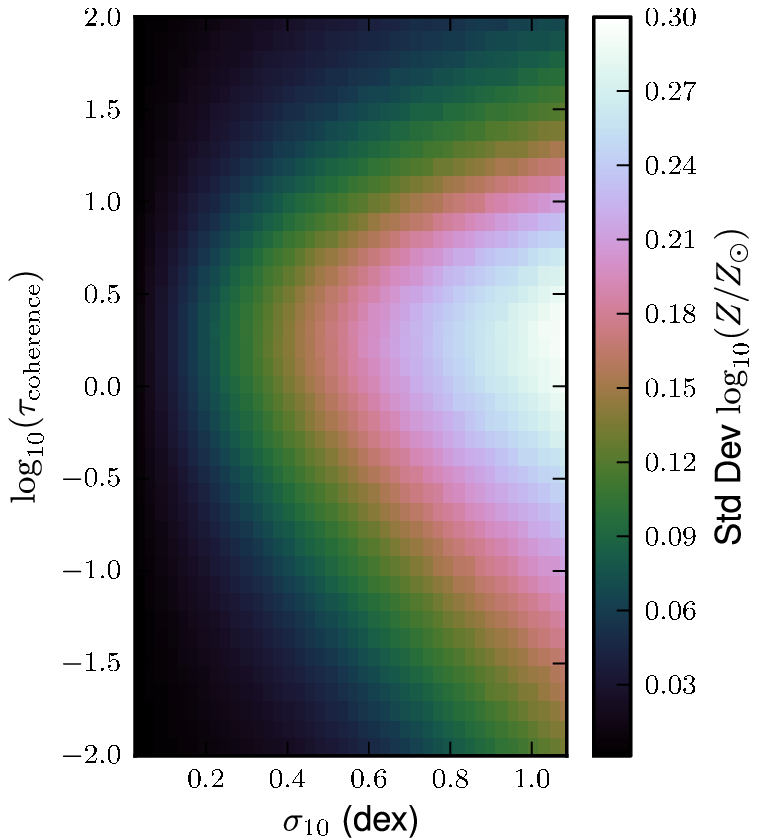}
\caption{The width of the mass-metallicity relation. Each pixel represents an ensemble of galaxies with fixed $\tau_c$ and $\sigma$, wherein the metallicity $Z$ was measured for each galaxy at a random time. The standard deviation of $\log Z/Z_\odot$ in each bin is plotted above. Longer coherence times (upwards in the plot) allow the galaxies to equilibrate so that $Z$ approaches $Z_{eq}$ (a constant value regardless of the accretion rate). Coherence times shorter than the mass loss timescale, i.e. $\log_{10} \tau_c \la 1$, lead to a reduction in the scatter roughly in proportion to $\tau_c$ -- individual draws from the accretion rate matter increasingly less. Larger intrinsic scatters (rightward in the plot) negate this effect by making large accretion events typical.}
\label{fig:sigZ}
\end{figure}

As in the previous section, we can construct a recursion relation by setting $\Delta=\tau_c$ in equation \ref{eq:Zdt}. As with the mass loss rate, we see that the metallicity at the switches in accretion rate is given by a sum wherein the effects of long-past accretion are exponentially suppressed,
\begin{eqnarray}
\label{eq:Zdk}
Z^\dagger_{k} &=& Z^\dagger_0 \prod_{i=0}^{k-1} \frac{r_i}{r_i+e^{\tau_c}-1} \\ \nonumber
& &+ \sum_{i=0}^{k-1} \frac{e^{\tau_c}-1 + \tau_c(r_i-1)}{e^{\tau_c}-1 + r_i} \prod_{j=i}^{k-1} \frac{r_j}{e^{\tau_c}-1 + r_j}
\end{eqnarray}
Since $r_k>0$ and $e^{\tau_c}>1$, each factor in both of the products is guaranteed to be between 0 and 1 -- for a median value of $r_k=1$ each factor becomes $e^{-\tau_c}$. In that sense this equation is very similar to equation \ref{eq:psik} for $\Psi_k$. We show how $Z^\dagger$ is constructed in figure \ref{fig:ConstructZ} -- in addition to the full distributions of $Z^\dagger$ and $Z^\dagger_k$, we show the distributions for galaxies with $\Psi$ in the top and bottom quartiles of galaxies with those values of $\sigma$ and $\tau_c$.  As in figure \ref{fig:ConstructPsi}, previous terms in the sum recede into irrelevancy over the course of a few star formation times. There are, however, crucial differences between the construction of $Z^\dagger_k$ and $\Psi_k$. To construct $\Psi_k$, independent lognormal variables were added together to get another quantity which was roughly lognormal. Here on the other hand, even though an individual $r_i$ is indeed roughly lognormally distributed (since $r_k=\Psi_k/e^{\sigma x_k}$ and $\Psi_k$ is roughly lognormal), it is not the $r_i$'s being summed, but rather a more complicated function of $r_i$ and $\tau_c$. Moreover, the $r_i$ are not independent of each other. Perhaps worst of all, since $Z^\dagger$ does not monotonically approach its equilibrium value each time the accretion rate changes, but rather increases to $Z^\dagger_\mathrm{extreme}$ before returning to $Z=Z_{eq}$, $Z^\dagger_k$ becomes a bad approximation of the full distribution of $Z^\dagger$ when $\tau_c \ga 1$ (though eventually, for $\tau_c \gg 1$, the probability distributions of $Z^\dagger$ and $Z^\dagger_k$ both approach $\delta(Z^\dagger-1)$).

\begin{figure}
\includegraphics[width=9cm]{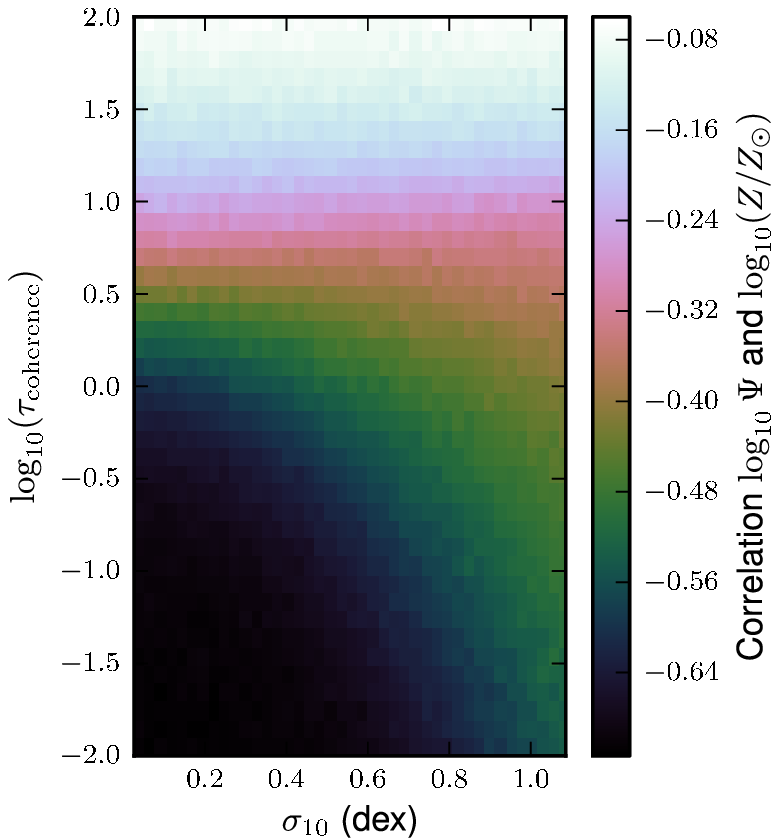}
\caption{The correlation between $\Psi$ and $Z/Z_\odot$. For small scatters and rapid variability in the accretion rate, the star formation rate and metallicity are substantially anti-correlated. Increasing the coherence time allows galaxies to return to their equilibrium $Z$ regardless of the accretion rate, wiping out the anti-correlation. }
\label{fig:corrSFZ}
\end{figure}

Despite these difficulties, we can see that $p(Z^\dagger) \approx p(Z^\dagger_k)$ for shorter accretion times, and we can also begin to see that $\Psi$ and $Z^\dagger$ are anti-correlated. Galaxies which have $\Psi$ in the top quartile of their ensemble tend to have lower metallicities. This is a very simple physical effect, namely the competition between dilution of metals by new infall and pollution of metals by stellar evolution. If the galaxy is burning through excess gas from previous accretion events ($r_k>1$), the metallicity will increase as the gas reservoir is polluted, whereas if the galaxy is accreting more gas than it is losing ($r_k<1$), the star formation rate is slow enough that new gas is added faster than metals can be produced to pollute it. Using the same Monte Carlo simulations used to produce figure \ref{fig:sigSF}, we can explicitly show the correlation between $\log \Psi$ and $\log Z$ (figure \ref{fig:corrSFZ}), defined in general according to 
\begin{equation}
Corr(X,Y) = \frac{\sum_i (X_i-\bar{X})(Y_i-\bar{Y}) }{ \sqrt{\left(\sum_i (X_i - \bar{X})^2 \right) \left(\sum_j (Y_j-\bar{Y})^2 \right)}},
\end{equation}
where each sum is over all galaxies in the ensemble and $\bar{X}$ indicates an average over the ensemble. The correlation is strongest in the `linear regime', namely small rapid perturbations. When the coherence time exceeds a few star formation times, the correlation disappears --  $Z^\dagger$ is always close to unity regardless of the accretion rate. The correlation also weakens as the intrinsic scatter in the accretion rate increases.

\begin{figure}
\includegraphics[width=9cm]{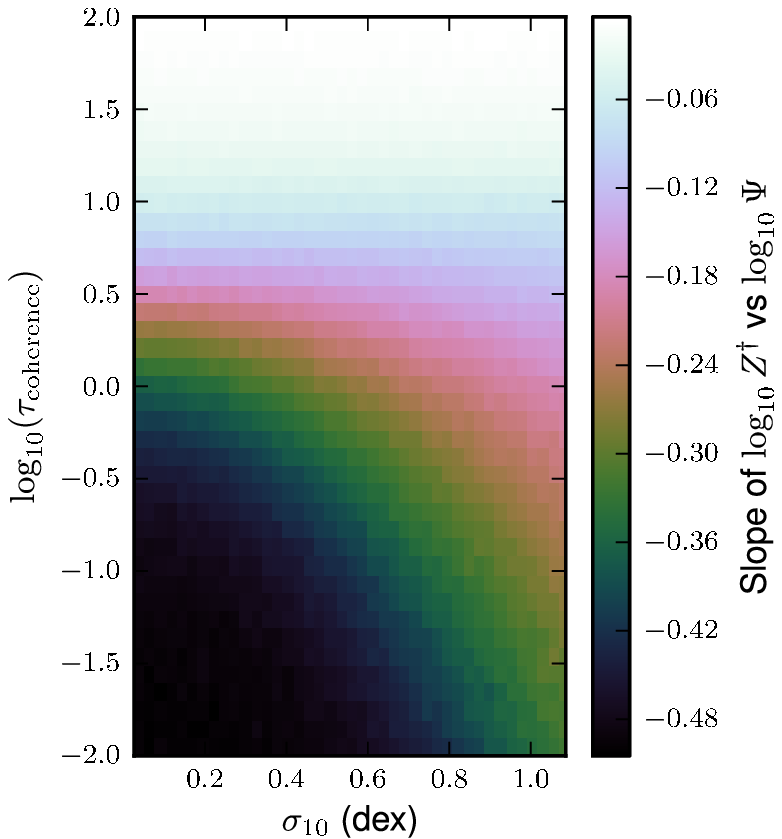}
\caption{The ``slope of the FMR''. For each $\tau_c$ and $\sigma$ we fit a linear model to the scatter plot of $\log_{10} \Psi$ vs $\log_{10} (Z/Z_\odot)$ of all galaxies drawn from the distribution, and plot the resulting slope here. We find uniformly that the slope is negative, though for the same reasons mentioned in the plot of the correlation between these two variables, the slope flattens for large values of $\tau_c$, and to a lesser degree, for large values of $\sigma$.}
\label{fig:fundSlope}
\end{figure}


\section{Construction of Galaxy Scaling Relations -- an initial guess}
\label{sec:firstguess}
The results we have derived thus far, namely the joint probability density of $\Psi$ and $Z^\dagger$, are applicable only at fixed values of $\mu$, $\sigma$, $t_\mathrm{coherence}$, $t_\mathrm{loss}$, $Z_{IGM}$, and $q$. We refer to an ensemble of galaxies with fixed values of these parameters as a simple (or stationary) galactic population (SGP).  As we saw in the previous sections, the only variables which affect the joint distribution of $\Psi$ and $Z^\dagger$ are $\sigma$ and $\tau_c = t_\mathrm{coherence}/t_\mathrm{loss}$. However, to compute physical quantities, i.e. the star formation rate, the metallicity, etc., one must specify the other variables.

The power of our approach using SGPs is that, to a reasonable approximation, the properties of a star-forming galaxy are set by a single parameter having to do with the size of the galaxy (e.g., stellar mass, halo mass, or K-band luminosity). One could therefore hope that, at a fixed stellar mass, the population of galaxies may be well-described by a single SGP. Essentially, all of the scalings of mass loading factor, accretion rate, etc. which account for the slope and zero-points of galaxy scaling relations would be taken out, leaving only the intrinsic scatter.

\begin{table*}
        \begin{minipage}{150mm}
        \caption{Important parameters used in this study}
        \label{ta:params}
        \begin{tabular}{cl}
        \hline
        Parameter &  Description \\  
        \hline
        &  {\bf The accretion process} \\
        $\mu$ & The log base $e$ of the median baryonic accretion rate. \\ 
        $\sigma$ & The log-normal scatter in the (DM and baryonic) and accretion rate \\
        $\Delta \omega$ & The scale-free time step taken to generate accretion histories \\
        $t_\mathrm{coherence}$ & The time over which the accretion rate is constant before a new random value is drawn. \\
        $\epsilon$ & The fraction of $f_b \dot{M}_{DM}$ which reaches the gas reservoir \\ 
        $f_\epsilon$ & A fixed fraction by which the baryonic accretion rate may be reduced -- fixed by 1st order scaling relations\\
        $Z_{IGM}$ &  Metallicity of infalling baryons, fixed at $2\times 10^{-4}$\\ \hline
        &  {\bf Star Formation } \\
        $\dot{M}_\mathrm{SF}$ & The star formation rate \\
        $M_g$ & The gas mass available to form stars \\
        $\eta$ &  The ratio of mass lost in galactic winds to stars formed -- fixed by 1st order scaling relations \\
        $t_{dep}$ & The depletion time, $M_g/\dot{M}_\mathrm{SF}$ -- fixed by 1st order scaling relations \\
        $f_R$  & The remnant fraction - fraction of mass not returned to the ISM from a simple stellar population, fixed at 0.54 \\
        $t_{loss}$ & The loss time, i.e. the timescale on which the reservoir is depleted, $t_{dep}/(\eta+f_R)$ \\
        $y$ &  Mass of metals returned to the ISM per mass locked in stellar remnants, fixed at 0.054 \\
        $\xi$ & Metallicity enhancement of galactic winds, fixed at 0 (perfect mixing) \\
        $q$ & A combination of $y$, $f_R$, $\xi$, and $\eta$ we call the effective yield \\ \hline 
        & {\bf Stationary Galactic Population (SGP) Parameters} \\ 
        $\tau_c$ & $t_\mathrm{coherence}/t_{loss}$ -- an input to the SGP \\
        $\sigma_{10}$ & This is the same $\sigma$ as above, but divided by $\ln 10$ -- an input to the SGP \\
        $\Psi$ & The ratio of the mass loss rate ($M_g/t_{loss}$) to the median accretion rate $e^\mu$ -- an output of the SGP \\
        $ Z^\dagger$ & The metallicity of the gas reservoir, subtracting out $Z_{IGM}$ and normalizing to $q$ -- an output of the SGP \\
        $x_k$ & For a given galaxy, the $k$th draw from a standard gaussian, which sets the accretion rate. \\ \hline
        & {\bf Other sources of scatter (see especially section \ref{sec:alternatives})} \\
        $\sigma_{\log_{10} M_*}$ & Log-normal scatter added to $M_*(M_h)$, assumed uncorrelated with anything else \\
        $\sigma_{\log_{10} t_{dep}}$ & Log-normal scatter added to the depletion time \\
        $\sigma_{\log_{10} \eta}$ & Log-normal scatter added to the mass loading factor $\eta$ \\
        $\sigma_{\mu}$ & Gaussian scatter in $\mu$. This is like including a stochastic component whose coherence time $\gg t_{loss}$.\\ \hline
        \end{tabular}
\end{minipage}
\end{table*}

Here we make an educated guess as to how to map the dimensionless model presented in the first few sections of this paper to observable galaxy scaling relations. Although alternative assumptions may be preferred by other practitioners, we hope this exercise will be at least illustrative. In section \ref{sec:accr} we use results from N-body-only dark matter simulations to make guesses for the parameters of the accretion process: $\mu$, $\sigma$, and $t_\mathrm{coherence}$. Each of these parameters thereby has a predicted scaling with halo mass and redshift. In this procedure we leave one free parameter $f_\epsilon$, a constant less than unity, to describe how much the accretion rate is reduced below this  guess. We then adopt the assumptions that $y=0.054$, $f_R=0.54$ (appropriate for a \citet{chabrier2005} IMF with yields from solar metallicity stars \citet{maeder1992} -- see appendix A of \citet{krumholz2012}), $Z_{IGM} \ll 0.02$, and $\xi=0$. To fully specify the SGP model, the only remaining parameters are $\eta$, $t_{dep}$, and $f_\epsilon$. In section \ref{sec:firstOrder}, we adopt a value of $f_\epsilon$ and powerlaw scalings of $t_{dep}$ and $\eta$ with halo mass such that we fit three galaxy scaling relations: $M_*$ vs. SFR, Z, and $M_g/M_*$. To do so we need to assign a value of $M_*$ to each halo mass, for which we take the \citet{behroozi2013,behroozi2013-a} relations with a fixed log-normal scatter of 0.19, consistent with various observational constraints \citep{behroozi2013,reddick2013}. 

With all of these parameters specified as a function of halo mass and redshift, we can construct synthetic versions of these galaxy scaling relations, including their intrinsic scatter, and the higher-order fundamental metallicity relation, $Z(M_*,\rm{SFR})$. Section \ref{sec:firstScatter} describes how we take the synthetic relations and extract three higher-order quantities which we will use to constrain our model: the scatter in the MS, the scatter in the MZR, and the ``slope'' of the FMR, namely the logarithmic derivative of the metallicity with respect to the star formation rate at fixed stellar mass. These three quantities can then be compared directly to observations, which allows us to rule out our initial guess, and non-trivially constrain $\sigma$, $t_\mathrm{coherence}$, and other potential sources of scatter (section \ref{sec:alternatives}). The parameters used throughout the paper are summarized in table \ref{ta:params}.



\subsection{Baryonic Accretion}
\label{sec:accr}
As is often the case for modellers of galaxies, we will begin with the dark matter. For the purposes of this simple model, we will rely on the EPS-like \citep{press1974, sheth1999} formalism presented in \citet{neistein2008} and \citet{neistein2010}. From this we will derive approximations for $\mu$, $\sigma$, and $t_\mathrm{coherence}$. For a WMAP5 cosmology, \citet{neistein2010} can fit the cumulative mass function of halos found in an N-body simulation if they construct halo accretion histories according to
\begin{equation}
\label{eq:dMh}
\Delta S = \exp(\mu_p + x \sigma_p)
\end{equation}
where $x$ is a standard normal drawn at a fixed interval $\Delta \omega$, and
\begin{equation}
\label{eq:mup}
\mu_p = (0.132 \log_{10} S + 2.404)\log_{10}(\Delta \omega) + 0.585 \log_{10} S  - 0.436
\end{equation}
\begin{equation}
\label{eq:sigp}
\sigma_p = (-0.333 \log_{10} S  -0.321)\log_{10}(\Delta \omega) +  0.0807 \log_{10} S +  0.622
\end{equation}
In these equations, $S$ is a measure of the amplitude of the dark matter power spectrum, and $\omega$ is a measure of time, similar to redshift.

Note that in general $\mu_p \ne \mu$ and $\sigma_p \ne \sigma$ because these refer to the mean and scatter in $S$-$\omega$, rather than $M$-$t$ space. To compute an estimate of the accretion rate, we take a grid of $M_h$ and $z$ -- for each $M_h$ we can compute $S$ and draw a large number of $\Delta S$ values, from which we can compute $\Delta M_h$, the change in halo mass. For each $z$ we can compute $\Delta t$, the time between the current redshift and a time $\Delta \omega$ earlier. We then approximate each baryonic accretion rate as
\begin{equation}
\dot{M}_\mathrm{ext} \approx f_b \epsilon \frac{\Delta M_h}{\Delta t}
\end{equation}
From our ensemble of $\Delta M_h$, we can approximate $\mu$ and $\sigma$ as the mean and standard deviation of $\ln\dot{M}_\mathrm{ext}$. We note that the distribution of $\dot{M}_\mathrm{ext}$ is not guaranteed to be log-normal, and this procedure produces distributions with non-zero skew and kurtosis in $\ln\dot{M}_\mathrm{ext}$. We neglect this non-gaussian component and approximate $\dot{M}_\mathrm{ext}$ as lognormal, keeping in mind that this is merely a guess at the true baryonic accretion rate. We take the efficiency factor to be
\begin{equation}
\epsilon = f_\epsilon  \min( 0.31 (M_h/10^{12} M_\odot)^{-0.25}  (1+z)^{0.38}, 1.0),
\end{equation}
where we leave $f_\epsilon < 1$ a free parameter to be fit in the next section. The remaining factors come from the fitting formula of \citet{faucher-giguere2011}, which accounts for the suppression of accretion in high-mass halos presumably due to hot virialized gas.

We estimate $t_\mathrm{coherence}$ by the fact that in the dark matter simulations, the merger trees become non-Markov for $\Delta \omega \la 0.5$, indicating that the accretion rates over time intervals shorter than that are correlated \citep{neistein2008}. We therefore use
\begin{equation}
t_\mathrm{coherence} \sim \left| \left(\frac{d\omega}{dz} \right)^{-1} \frac{dt}{dz} \Delta \omega \right|, 
\end{equation}
with $\Delta \omega = 0.5$. Obviously this is only a rough estimate, since a real accretion history is likely to have more structure in Fourier space than the single period we assume here. We encourage those with cosmological simulations to measure this quantity, both in dark-matter only and baryonic simulations. Another plausible value for $t_\mathrm{coherence}$ might be the dynamical time of the halo, or some other timescale related to the baryon cycle. This choice might be appropriate if the primary supply of gas is re-accreting winds.

\subsection{Fitting the first-order relations}
\label{sec:firstOrder}

To compute $\tau_c$ and re-dimensionalize the SGP, we still need to know the mass-loss timescale, i.e. the depletion time and $\eta$, as well as $\xi$. These values are sufficiently uncertain that it is worth digressing to discuss how they may be chosen to fit the first-order galaxy scaling relations, namely the star-forming main sequence, the mass-metallicity relation, and the stellar mass-gas mass relation.

To compare our model with many galaxy scaling relations, computed as a function of stellar mass $M_*$, we must pick an $M_*$. Since our model is purely equilibrium-based, we have no way to compute integrated quantities like $M_*$ besides appealing to other empirical relations. We employ the \citet{behroozi2013,behroozi2013-a} model of the relation between stellar mass and halo mass, including a scatter in stellar mass at fixed halo mass of 0.19 dex \citep{reddick2013}. Another common approach is to simply use $M_*$ rather than $M_h$ \citep{lilly2013} as the parameter by which to scale the SGP parameters ($\eta$, $\xi$, etc.)

In a standard equilibrium model \citep{dave2012}, the `center' of a galaxy scaling relation may be determined by setting time-derivatives to zero. For instance, setting $dM_g/dt = 0$ yields an equilibrium star formation rate
\begin{equation}
\label{eq:psieq}
\dot{M}_{\mathrm{SF},eq} = \dot{M}_\mathrm{ext}/(\eta+f_R).
\end{equation}
Immediately we can see the critical importance of two unknown pieces of physics -- anything which displaces the baryonic accretion rate $\dot{M}_\mathrm{ext}$ away from the naive estimate $f_b \dot{M}_{DM}$ (e.g. preheating, halo quenching, AGN heating), and anything which removes gas from the star-forming gas reservoir of the galaxy (supernovae, radiation pressure, cosmic rays, or direct AGN). Typically the former are invoked at high mass and the latter at low mass.

Interestingly, the depletion timescale does not enter into $\dot{M}_{\mathrm{SF},eq}$. This is because the gas mass is free to adjust to whatever it needs to be so that inflowing gas is balanced by sinks for the gas -- old stellar remnants and galactic outflows. For instance, a longer $t_{dep}$ would simply mean a larger gas mass would be required to reach the same equilibrium between inflows and sinks.
\begin{equation}
\label{eq:mgeq}
M_{g,eq} = \dot{M}_{\mathrm{SF},eq} t_{dep} = \dot{M}_\mathrm{ext} t_\mathrm{loss}
\end{equation}
Thus we can see that star formation rates can only be affected by physics which alters the star formation timescale \citep[e.g. H$_2$ regulation][]{krumholz2012} to the degree that the galaxies are out of equilibrium. These physical considerations do, however, affect the equilibrium size of the gas reservoir.

If we make the additional restriction that $dZ/dt = 0$, we arrive at the equilibrium metallicity
\begin{equation}
\label{eq:Zeq}
Z_{eq} = Z_{IGM} + q = Z_{IGM} + \frac{y f_R}{f_R+\eta}\left(1 - \frac{\eta\xi}{\max(1-f_R,\eta)}\right).
\end{equation}
The metallicity is a particularly powerful probe of feedback physics because $Z_{eq}$ is independent of the accretion rate, meaning that ``preventive feedback'' or any other considerations which affect $\dot{M}_\mathrm{ext}$ do not affect the first-order mass-metallicity relation (MZR) -- essentially all that matters are $\eta$, the mass loading factor, and $\xi$ or some other measure of the mixing between ejecta and the ISM. The drawback of using the MZR is that it is bedeviled by large systematic uncertainties in converting characteristics of metal emission lines into actual gas-phase metal abundances \citep{kewley2008}. 

Given these equilibrium relations, we attempt to find powerlaw scalings of $\eta$ and $t_{dep}$, and a constant value of $f_\epsilon$, which will roughly fit the observed galaxy scaling relations. This leads us to take $f_\epsilon = 0.5$, $\eta = (M_h/10^{12} M_\odot)^{-2/3}$, and $t_{dep} = $ 3 Gyr $(M_h/10^{12} M_\odot)^{-1/2}$, as described below. These fits were done by hand, which is reasonable given the uncertainties in the mean relations.

To approximately fit the centers of the observed $z=0$ main sequence, we had to use a steep scaling of the mass loading factor with halo mass, namely $\eta = (M_h/10^{12} M_\odot)^{-2/3}$. This is a result of the short cooling times and/or cold streams in low-mass halos which efficiently supply cold gas at a rate proportional to the dark matter accretion rate, which is much larger than the observed star formation rate in these galaxies. That is, $\epsilon$ is large for low-mass galaxies in our initial guess. If low-mass galaxies are in statistical equilibrium, these large values of $\eta$ are necessary. It is possible that some other physical process is decreasing $\epsilon$ at these masses, or that the galaxies are not in equilibrium \citep{krumholz2012,kuhlen2012,kuhlen2013}. In the latter scenario, $\eta$ may vary quite weakly, leaving the mass loss timescales at low masses to be comparable to the depletion times, themselves comparable to or much longer than the age of the universe. We discuss this possibility further in section \ref{sec:applicability}, but for now we adopt $\eta \propto M_h^{-2/3}$.

\begin{figure}
\includegraphics[width=9 cm]{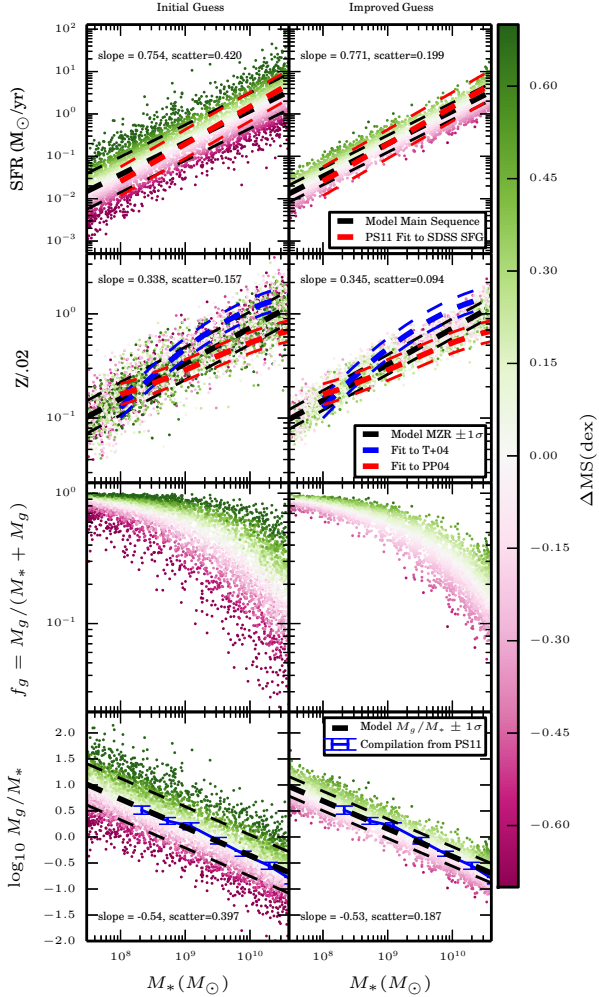}
\caption{Scaling relations. We compare observed galaxy scaling relations and their scatters (colored lines) with scaling relations (black lines) calculated by drawing galaxies (colored points) from a sequence of SGPs using reasonable guesses of how the SGP parameters scale with halo mass. The left panel is our initial guess, while the right panel shows an improved guess which better fits the observational constraints on the scatter in these relations. In particular, the initial guess yields a model MS with intrinsic width wider than the observed scatter, while the improved guess has a much small intrinsic scatter. The colors indicate offset from the Main Sequence, namely the black dashed line fit in the top panel. PS11 refers to \citet{peeples2011}, and T+04 and PP04 refer to metallicity callibrations used by \citet{tremonti2004} and \citet{pettini2004}. The $M_g/M_*$ data come from \citet{mcgaugh2005,leroy2008,garcia-appadoo2009}.}.
\label{fig:firstguess}
\end{figure}

We do not fit the MZR particularly well in this simple model (see figure \ref{fig:firstguess}). That is, most calibrations of the MZR are not well-described by the powerlaw in $M_h$ we assume here. This is not extremely concerning given the systematic uncertainties in the observations. The scaling of $\eta$ needed to fit the star-forming main sequence does leave the MZR in the right neighborhood, without adjusting the fiducial values of $Z_{IGM} = 2 \times 10^{-4} \approx Z_\odot/100$, $\xi=0$, or $f_\epsilon$ and $\eta$.

In contrast, the gas fraction data are fit remarkably well by our fiducial scalings. Given the reasonable fit of the SFR-$M_*$ relation, we were left with one parameter to vary to fit $M_g/M_* - M_*$, namely $t_{dep}$. Despite the (lack of) trends from the GASS \citep{schiminovich2010} and COLD GASS \citep{saintonge2011} surveys in the total gas depletion time with mass, we find that we need $t_{dep} = $3 Gyr $(M_h/10^{12}M_\odot)^{-1/2}$. Since $\eta$ scales even more steeply, the mass loss timescale actually shortens with decreasing halo mass in this scenario, scaling as $t_\mathrm{loss} \propto M_h^{1/6}$.

\begin{table}
        \begin{minipage}{78mm}
        \caption{The two reference models we use to generate synthetic galaxy scaling relations and FMRs}
        \label{ta:guesses}
        \begin{tabular}{ccc}
        \hline
        & Initial Guess\footnote{Based on N-body simulations} &  Improved Guess\footnote{Adjusted to fit both the conservative observational constraints, and the much tighter $-2 < \partial \log Z/\partial \log \dot{M}_\mathrm{SF} < -1$  } \\  
        \hline
        $\sigma$ (dex) & 0.45 & 0.225 \\
        $\Delta \omega$ & 0.5 & 0.2 \\
        $\sigma_{\log_{10} M_*}$ (dex) & 0.19 & 0.07 \\
        $\sigma_{\log_{10} t_{dep}}$ (dex) & 0 & 0 \\
        $\sigma_{\log_{10} \eta}$ (dex) & 0 & 0 \\
        $\sigma_{\mu}$ (dex)& 0 & 0.08\\ \hline
         \end{tabular}
\end{minipage}
\end{table} 

The result of applying these scalings is shown in figure \ref{fig:firstguess}. For each $M_h$ below $10^{12.3} M_\odot$ provided in the \citet{behroozi2013,behroozi2013-a} $M_*(M_h)$ relation, we apply the scalings above to compute the necessary physical parameters to specify the SGP, draw 300 samples from the SGP, and assign each one a stellar mass according to $M_*(M_h)$, plus a fixed scatter, $\sigma_{\log_{10} M_*} =$ 0.19 dex. The resultant population of galaxies, as designed, fits various observational constraints at $z=0$, represented in each figure by the various colored lines (which include representative scatters) from \citet{peeples2011}. The black lines show power law fits and the computed $\pm 1 \sigma$ scatter of the population. The observational fits are not necessarily precise, and in particular the mass-metallicity relation famously has many different fits depending on which calibration is used \citep{kewley2008}. The two columns in figure \ref{fig:firstguess} have the same scalings of $t_\mathrm{dep}$, $\eta$, and $q$, but different values of $\sigma$, $\Delta\omega$, and $\sigma_{\log_{10} M_*}$. The left column shows the initial guess discussed here and in the previous section, while the right column has values of the accretion process parameters more in line with observations (the ``Improved Guess'' model -- see table \ref{ta:guesses} and sections \ref{sec:firstScatter} and \ref{sec:alternatives}).

It is worth emphasizing that fitting or not fitting the observed relations should be construed neither as success nor failure for our model -- the equilibrium relations (equations \ref{eq:psieq}, \ref{eq:mgeq}, \ref{eq:Zeq}) have enough free parameters that fitting the observations is not challenging. However it is encouraging that relatively little tuning was required for reasonable fits, and that we end up with models with a reasonable physical basis (energy-driven winds for $\eta\propto M_h^{-2/3}$, long depletion times at low masses owing to low H$_2$ fractions, $\eta$ of order unity and $t_{dep}$ of order 3 Gyr at high masses).

\subsection{Information in the scatter}
\label{sec:firstScatter}
Now that we have a reasonable fit for the redshift zero median relations, we can return to our main goal -- understanding the higher-order relationships in the data. In particular, we would like to understand the scatter in the MS and the MZR, and the (negative) slope implied by the FMR of metallicity with respect to star formation rate at fixed stellar mass. Now that we have synthetic data, we can follow a simple procedure to fit a synthetic MS, MZR, and FMR. For the former two, we simply fit a linear model with least-squares regression to (the log base 10) of SFR vs $M_*$ and Z vs $M_*$. For each synthetic galaxy, we can then subtract off the linear fit for SFR or Z at that galaxy's $M_*$ to find its residual. Finally, the scatter is calculated as the sample standard deviation of the residuals. 

Meanwhile, we fit a synthetic FMR both with a linear model 
\begin{equation}
\log_{10} Z = b_0 + b_1 \log_{10} M_* + b_2 \log_{10}  \dot{M}_\mathrm{SF}. 
\end{equation}
and, as is common practice, a quadratic model\footnote{Note that the quadratic fits reported in the literature are typically fits to $12 + \log_{10}(O/\mathrm{H})$, whereas here we are fitting to the metallicity. This only makes a difference in the $a_0$ term, and not to the slope with respect to the star formation rate at fixed stellar mass, on which we will concentrate.}
\begin{equation}
\label{eq:quadratic}
\log_{10} Z = a_0 + a_1 m + a_2 s + a_3 m^2 + a_4 m s + a_5 s^2 
\end{equation}
Here $m=\log_{10}M_*/M_\odot - 10$ and $s = \log_{10} \dot{M}_\mathrm{SF}/(M_\odot\ \mathrm{yr}^{-1})$. From the linear fit, we can read off a value for $\partial \log_{10} Z / \partial \log_{10} \dot{M}_\mathrm{SF}$ at fixed $M_*$, namely $b_2$. For the quadratic fit, the slope is a function of both mass and $\dot{M}_\mathrm{SF}$.

For all three of these quantities -- the two scatters and the (logarithmic) slope of $Z$ vs $\dot{M}_\mathrm{SF}$ at fixed stellar mass, we can compare both to observations and to predictions from the full joint distribution of $\Psi$ and $Z^\dagger$ by a single SGP (figures \ref{fig:sigSF}, \ref{fig:sigZ}, and \ref{fig:fundSlope}). The scatters in the MS, MZR, and FMR are not identical to that in a SGP because the relationship between stellar and halo mass also has some scatter. Thus a population of galaxies at fixed $M_*$ represents a weighted sum of SGPs.

\begin{figure}
\includegraphics[width=9 cm]{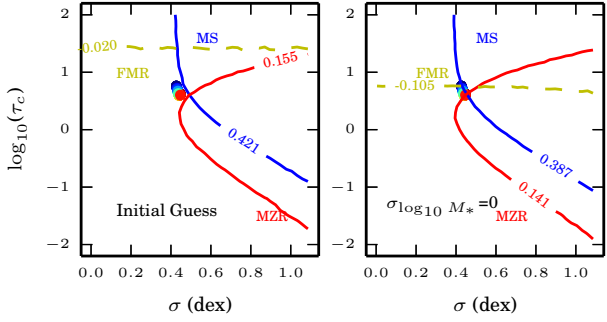}
\caption{Information in the scatter of the MS and MZR. Here we show contours extracted from the SGP predictions, namely the values of $\tau_c$ and $\sigma$ that we would have predicted given the synthetic scatters and FMR slope. The input values are shown as colored points near (.45,.8), where colors represent a wide range of halo mass. If we artificially reduce the scatter in $M_*$ at fixed halo mass (right panel), the SGP predictions recover the correct input values, but including the scatter (left panel) increases the synthetic scatter and flattens the FMR slope.}
\label{fig:contours}
\end{figure}

Thus there will be additional scatter in these synthetic observations, as compared to the SGP predictions. This is illustrated in figure \ref{fig:contours} -- we compare the input values of $\tau_c$ and $\sigma$ (the colored points, nearly on top of each other) to contours extracted from the heatmaps. In particular, given the synthetic scatters and FMR slope, we can read off from the SGP figures values of $\tau_c$ and $\sigma$ which are consistent with these synthetically-observed values. When we artificially set $\sigma_{\log_{10} M_*} = 0$, these contours converge at the input values, but including the scatter in $M_*$ flattens the FMR slope and increases the scatter in the MS and MZR, leaving a general region in $\tau_c$ - $\sigma$ space where the contours are close to each other, but no trivial way to recover the input values.

This at once shows both the promise and the difficulty of using observed second order relations to predict these parameters. The constraints are largely independent of each other, so one could hope to nontrivially constrain the acceptable values of $\sigma$ and $\tau_c$. However, additional sources of scatter not included in the simple dimensionless SGP predictions can make it difficult to recover the values of these parameters simply by reading off where the contours intersect in this diagram. 

Nonetheless, this diagram (figure \ref{fig:contours}) and the associated SGP predictions (figures \ref{fig:sigSF}, \ref{fig:sigZ}, and \ref{fig:fundSlope} ) provide some heuristic guidance. We see that in this parameter space, the input values of $\sigma$ and $\tau_c$ are nearly independent of halo mass. This means that the fiducial model would predict no change in the scatters of the MS or MZR, nor any change in the slope of the FMR, which is roughly consistent with observations. We also see that to reduce the synthetic scatter to below the observed scatter, $\la \pm 0.34$ dex \citep{whitaker2012} for the MS and $\la \pm 0.1$ dex \citep{kewley2008} (both of which may be regarded as upper limits on the intrinsic scatters), we could reduce the input accretion scatter, $\sigma$, or dramatically reduce the coherence time (and therefore $\tau_c$). Our ``Improved Guess'' model adjusts the initial guess to reduce the two scatters and steepen the FMR slope. In particular we reduce $\sigma$ by a factor of two and $\Delta\omega$ in our estimate of $t_\mathrm{coherence}$ to 0.25 (see next section).

\section{Discussion}
\label{sec:discussion}

In the previous section we set up a fiducial set of assumptions to map the stationary galactic populations of sections \ref{sec:ms} and \ref{sec:Z} into observable parameters. We were easily able to match the first-order relations, but our first guess produced scatters in our synthetic MS and MZR that were too large. In this section we examine in more detail the full range of SGP-based models that are consistent with the observed constraints on the scatters in the MS and MZR and the slope of the FMR, and the range of halo masses and redshifts over which SGP-based models are valid in principle.

\subsection{A more general model -- do all galaxies at a fixed $M_h$ correspond to one SGP?}
\label{sec:alternatives}
The analysis of SGPs in sections \ref{sec:ms} and \ref{sec:Z} explicitly assumes that a given galaxy has had the same values of $\mu$, $\sigma$, $t_\mathrm{loss}$, $t_\mathrm{coherence}$,$Z_{IGM}$, and $q$ for eternity. This is clearly false -- galaxies increase their mass over time, moving them along any presumed scaling relations in e.g. $t_{dep}$ or $\eta$, while other quantities likely depend explicitly on time, e.g. $\mu$ and $t_\mathrm{coherence}$. The statistical equilibrium model we have proposed here, and other simpler models, may still be successful in describing galaxies because these quantities plausibly vary slowly relative to the internal timescales of the galaxy, i.e. the loss time. Whereas the typical equilibrium model assumes this of the accretion rate, our model relaxes that particular assumption and allows the accretion rate to vary, possibly very quickly, relative to other timescales.

Our model was constructed with the goal of understanding the scatter in galaxy scaling relations by examining the role of a known (and significant) scatter in dark matter accretion rates among galaxies at a given mass. However, it is also plausible that the mass loading factor, the depletion time, or some other quantity may vary between galaxies near a fixed mass, or within a given galaxy on relatively short timescales. The former situation may be handled by our model by having multiple SGPs with different values of e.g. $\eta$ at the same mass. The latter situation cannot be handled by SGPs as we have formulated them.

The scenario in which scatter in $\eta$, $t_{dep}$, or $\mu$ is responsible for the scatter in galaxy scaling relations has several distinct predictions compared to the stochastic accretion model we have presented. In particular, the scatter, rather than being stochastic, would be constructed from several nearly parallel, slightly offset, equilibrium relations. One could likely find acceptable values for the scatter in the SGP parameters which reproduced the observed scatters, since each equilibrium relation depends on a different combination of the SGP parameters. Thus one might expect to be able to have $e^\mu$ vary by $\sim 0.34$ dex, and $Z_{IGM}$ to vary by $\sim 0.1$ dex. 

\begin{figure}
\includegraphics[width=9 cm]{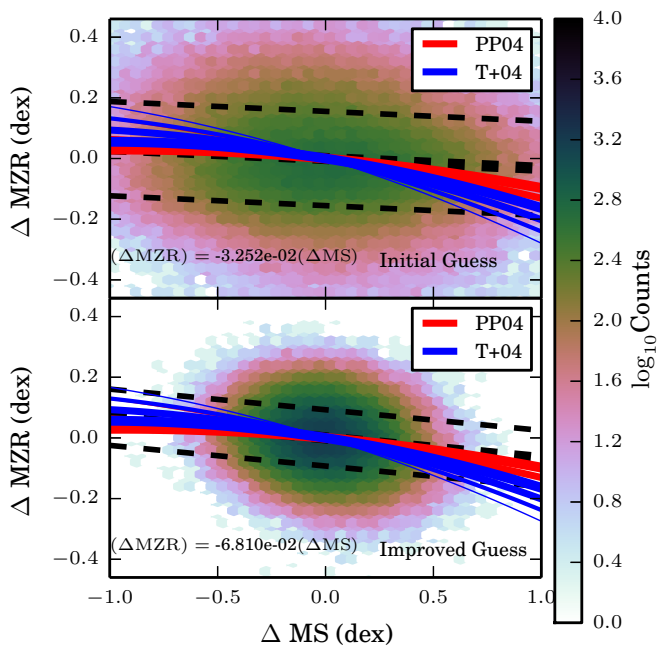}
\caption{Displacements from first-order scaling relations. Here we show the offset of galaxies, in both our initial guess and improved guess models, from fits to their Main Sequences and Mass-Metallicity Relations. The black lines show a linear fit to these data points, with a $1\sigma$. The colored lines show predictions of these quantities from two different fits to the $z=0$ FMR using different metallicity calibrations, with thicker lines corresponding to larger values of $M_*$. Here we can see the slight, but significant observed and predicted anti-correlation between star formation rate and metallicity. Note that to fill out this histogram we drew a sample of 100 times the number of galaxies shown in figure \ref{fig:firstguess}}
\label{fig:delta}
\end{figure}

This model would indeed produce, more or less, the observed scatters in the MS and MZR, but it would not account for the decreasing metallicity with increasing star formation rate at fixed stellar mass (see figure \ref{fig:delta}), i.e. what we call the `slope' of the FMR. In particular, the equilibrium metallicity is independent of the accretion rate, and the star formation rate is independent of $Z_{IGM}$ (at least in this simple model), so there would be no slope in the FMR. The situation gets even worse with a scatter in $\eta$, since both the equilibrium SFR and Z are inversely related to $\eta$, which would tend to create a positive slope in the FMR. Similarly, a scatter in $M_*$ at fixed halo mass tends to induce a positive slope -- at a given $M_*$, galaxies from higher halo mass and lower halo mass will be present, and because both the MS and MZR have positive slopes, the higher (lower) halo mass galaxies will have higher (lower) SFRs and Zs, again leading to a positive slope in the FMR.

\begin{figure}
\includegraphics[width=9 cm]{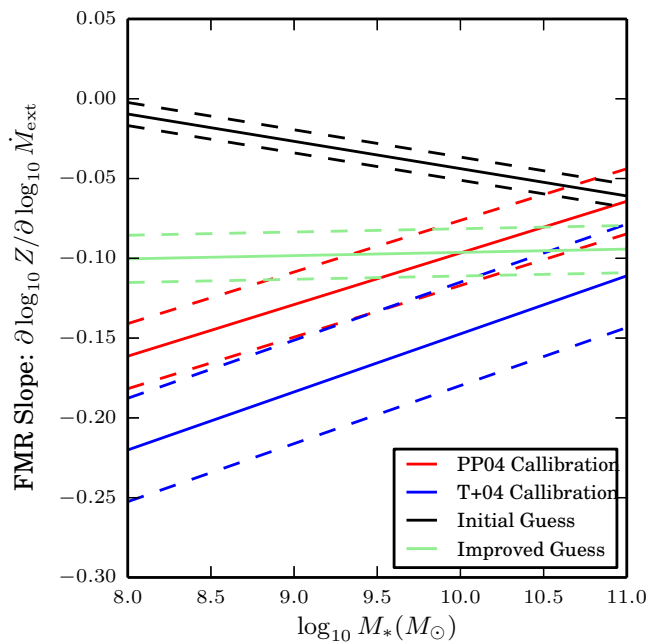}
\caption{Slope in the FMR. For both our improved and initial guesses, as well as the two fits to the $z=0$ FMR using different metallicity calibrations, we show the slope in the metallicity with respect to the star formation rate, as a function of stellar mass. Since the FMR fits depend explicitly on $\dot{M}_\mathrm{SF}^2$, we must also choose a star formation rate at which to evaluate this quantity. We choose the main sequence value at that mass (as determined by our fit to the main sequence of our initial guess model), which we plot as the solid lines, and $\pm 0.3$ dex, the dashed lines. The two different calibrations predict substantially different values, so to be maximally conservative we have chosen to interpret the observational constraint as $\partial \log Z/\partial \log \dot{M}_\mathrm{SF} < 0 $, a fact on which both observed relations agree. }
\label{fig:slope}
\end{figure}

Given the difficulty in obtaining a negative FMR by adding scatters in the parameters which enter the equilibrium relations, compared with the natural way the negative slope arises in our statistical equilibrium model, via a time-varying accretion rate, it certainly seems that no alternative model is needed. In fact, by enforcing the requirement that $\partial \log Z/\partial \log \dot{M}_\mathrm{SF} <0$, we may be able to obtain limits on the scatter in parameters ($M_*$, $\eta$) which tend to make the slope in the FMR positive.

To accomplish this, we set up a 6-dimensional grid of models. Each point in the grid corresponds to a choice of $\sigma$, $\log_{10} \Delta \omega$, scatter in $M_*$, scatter in $t_\mathrm{dep}$, scatter in $\eta$, and scatter in $e^\mu$. For each point in the grid, we simulate a full set of galaxies -- 200 per value of $M_h < 10^{12.3} M_\odot$, and compute the scatter in the MS and MZR and the slope in Z vs SFR at fixed $M_*$. We then compare each of these pieces of information to the observations in a maximally conservative way. We treat the observed scatters in the MS and MZR as upper limits on the intrinsic scatters, and make no assumptions about the purely observational scatter. Although there is an observationally known value of the slope of Z vs SFR at fixed $M_*$, we make no strong assumptions about the probability distribution function of that parameter -- we merely require that it is negative (see figure \ref{fig:slope}). Thus for each point in the grid, we can say how many constraints that model violates: 0, 1, 2, or 3.

\begin{figure}
\includegraphics[width=9 cm]{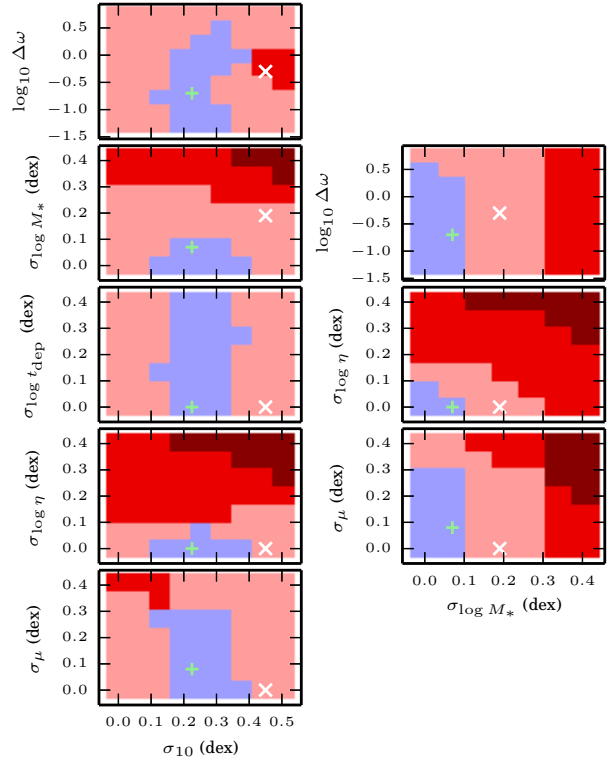}
\caption{The set of parameters conservatively allowed by the observations. In addition to the parameters of the accretion process ($\sigma_{10}$, and $\Delta\omega$) we include a variable log-normal scatter in $\eta$, $M_*$, $t_\mathrm{dep}$, and the median accretion rate $e^\mu$. These log-normal scatters have medians equal to the values used in section \ref{sec:firstOrder} to fit the first order relations. Blue pixels indicate that at least one model with that pair of parameters is consistent with the data. Each darker shade of red means the model which violates the fewest constraints for that pair of parameters violates one more constraint, up to all three. The white cross shows our initial guess, while the green '+' shows our ``Improved Guess''.}
\label{fig:scores}
\end{figure}

In figure \ref{fig:scores}, we project these scores, again in a maximally conservative way. For each point in the 2-d projection, we look up all models in the full 6-d space which have the 2 values under consideration in our projection, and we find the model which violated the fewest constraints. Thus the figure shows the minimum number of constraints violated by {\it any} model with that combination of values. If any model with those coordinates is allowed by the constraints, the pixel is shown in blue. Each darker shade of red means every model with those coordinates violated at least one more constraint, up to all 3. 

This exercise demonstrates that even this conservative interpretation of the observed scatters as upper limits, combined with the weak requirement that Z decrease with increasing SFR at fixed stellar mass, yields non-trivial constraints on the parameters. In particular, $\sigma \la 0.35$ dex, smaller than in our fiducial model. This may point to a smoothing out of the baryonic accretion rate relative to the scatter in dark matter accretion rates implied by the \citet{neistein2010} formula. Perhaps even more interesting is that there is a minimum $\sigma$ implied by our observational restrictions, $\sigma \ga 0.1$ dex, which comes from the requirement that the FMR have a negative slope. In particular, if $\sigma$ is too small, the subtle feature in the $M_*(M_h)$ relation causes both the star formation rate and metallicity of galaxies with $M_* \sim 10^9 M_\odot$ to be higher than the MS and MZR, and galaxies at other masses to be below those relations, generating a positive correlation between SFR and Z.

The scatter in $M_*$ for a given SGP must be $\la 0.15$ dex. This is a bit at odds with the observational constraints from \citep{reddick2013} ($0.20 \pm 0.03$ dex at fixed maximum circular velocity), although we note that our constraint is on scatter in $M_*$ that is {\it uncorrelated} with everything else in the SGP, whereas in reality it is quite plausible (and in fact predicted by the SGP -- see appendix \ref{app:mc}) that, at a fixed halo mass, $M_*$ is correlated with both SFR and Z. 

Another interesting constraint is that the scatter in $\eta$ must be $\la 0.1$ dex. This is surprisingly small, considering the great deal of theoretical uncertainty as to the actual values and scalings of $\eta$ in the first place. In our models, this comes from the aforementioned effect that in the equilibrium relations, both $Z$ and $\dot{M}_\mathrm{SF}$ are inversely related to $\eta$, so scatter in $\eta$ tends to reverse the negative slope in the FMR. 

Unsurprisingly, there is virtually no constraint on the scatter in $t_{dep}$. This is simply because the equilibrium relations for $\dot{M}_\mathrm{SF}$ and $Z$ are independent of $t_\mathrm{dep}$ -- to constrain this scatter one would need constraints on the scatter in the $M_*$ - $M_g/M_*$ relation, although if such galaxies also had SFR measurements, they would have directly measured depletion times anyway. There is also relatively little constraint on scatter in the median accretion rate, $e^\mu$. Essentially this is because, in the equilibrium relations, only the star formation rate is affected by this scatter, so as long as the scatter in $e^\mu$ is smaller than the scatter in the main sequence, there is no problem.

\begin{figure}
\includegraphics[width=9 cm]{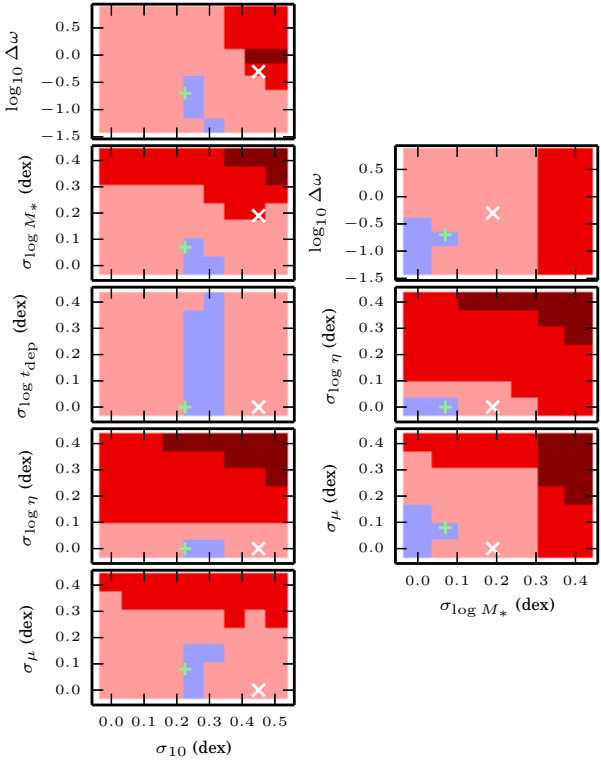}
\caption{A plausible set of constraints on model parameters. Here we make the plausible but uncertain assumption that the FMR slope, $\partial \log Z/\partial \log \dot{M}_\mathrm{SF}$, is between $-2$ and $-1$, in addition to the constraints on the widths of the MS and MZR. As one might expect, narrowing the allowed range of FMR slopes dramatically reduces the allowed regions of parameter space. One should not take these regions to be genuine constraints, but rather to demonstrate the power of the slope of the FMR in constraining these parameters.}
\label{fig:scores_low}
\end{figure}

With these constraints in mind, we have altered our initial guess, simply by reducing $\sigma$, $\Delta\omega$, and $\sigma_{\log_{10}M_*}$, and slightly increasing $\sigma_\mu$ (see table \ref{ta:guesses}). We label this model the ``Improved Guess'' model. Purely for demonstration, we have also computed scores for the grid of models where we include not only an upper limit on the scatters, but also a much narrower range of acceptable slopes of the FMR, namely $-2 < \partial \log Z/\partial \log \dot{M}_\mathrm{SF} < -1$. With these stronger restrictions, we get the projections shown in figure \ref{fig:scores_low}. Our ``Improved Guess'' model is engineered to adhere to this much stronger constraint, though there are plenty of models which would be ruled out by this strict scoring that are still consistent with the observations. Unsurprisingly the stronger constraint dramatically narrows the range of acceptable models in most of the projections. Particularly striking is that allowed range of $\Delta\omega \sim \Delta z$, the interval in redshift over which galaxies have constant accretion rates in our model, is narrowed substantially from $\Delta\omega \la 3$ to $\Delta\omega \la 0.4$

\subsection{Domain of applicability}
\label{sec:applicability}
Under what circumstances might a real population of galaxies be in statistical equilibrium? We know that for a constant accretion rate, the star formation rate and the metallicity will equilibrate on the mass loss timescale. A standard equilibrium model therefore requires that $t_\mathrm{loss}$ be much less than the timescale on which any parameter entering into the equilibrium relations, namely $q$ (i.e., $\eta$, $\xi$, and $f_R$), $Z_{IGM}$, $t_{dep}$ and $\dot{M}_\mathrm{ext}$. The success of these models in understanding the first-order trends in galaxy scaling relations suggests that these requirements, while seemingly numerous, are at least marginally satisfied.

Our statistical equilibrium model relaxes one of these restrictions by splitting $\dot{M}_\mathrm{ext}$ into a (hopefully) slowly-evolving median $e^\mu$ and a (potentially) rapidly varying stochastic component $e^{\sigma x(t)}$. Our formulation of this component introduces two timescales, $t_\mathrm{coherence}$ -- the time between new draws from the lognormal distribution, and $\sigma t_\mathrm{loss}$ -- the time for a 1-$\sigma$ accretion event to be forgotten by the galaxy.

Figures \ref{fig:ConstructPsi} and \ref{fig:ConstructZ} show graphically the exponential suppression of old draws of the accretion distribution in their influence on the full distribution of $\Psi$ and $Z^\dagger$. In logarithmic space, the separation between the centers of the distribution of each draw is just $\tau_c = t_\mathrm{coherence}/t_{loss}$, while the width of each distribution is $\sigma$. When $\sigma \la \tau_c$, the distributions are well-separated, and we conclude that galaxies may be in statistical equilibrium so long as $t_\mathrm{loss}$ is appreciably less than the timescale on which any parameters of the SGP change (explicitly $\mu$, $\sigma$, $t_\mathrm{coherence}$, $t_\mathrm{loss}$, $Z_{IGM}$, and $q$). Note that $t_\mathrm{coherence}$ itself may well be shorter than or comparable to $t_\mathrm{loss}$.

When $\sigma \ga \tau_c$, the contributions from previous draws begin to matter significantly for the distribution of $\Psi$. In this case, the number of draws which are important increase from $\sim 1$ to $\sim \sigma/\tau_c$, so rather than $t_\mathrm{loss}$ being short, we need $\sigma t_\mathrm{loss}$ to be short. At least in our initial guess, shown in the previous section, this is a minor effect since $\sigma \sim 1$. Therefore the region of $M_h$-z space where the statistical equilibrium model is valid should be comparable to the region where an ordinary equilibrium model is capable of reproducing the first-order galactic scaling relations, which in turn is set by the scaling of $t_\mathrm{loss}$ with halo mass and redshift.

\begin{figure}
\includegraphics[width=9 cm]{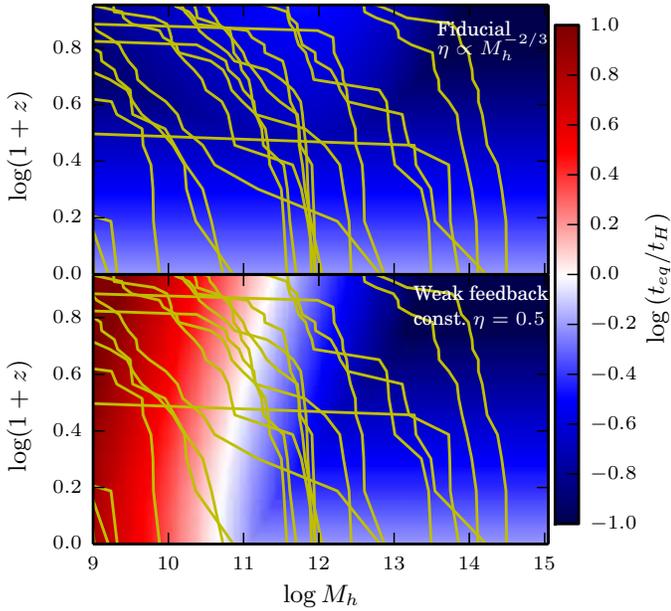}
\caption{The validity of the statistical equilibrium model. Here we show the ratio of $t_{eq}$ to the instantaneous Hubble time, as a function of halo mass and redshift. The bluer the color, the more satisfied the condition that $t_{eq} < t_H$, required for the statistical equilibrium model to be valid. For reference, we overplot the trajectories of 20 haloes as they grow stochastically, as calculated by equations \ref{eq:dMh}, \ref{eq:mup}, and \ref{eq:sigp} with $\Delta\omega = 0.5$.}
\label{fig:valid}
\end{figure}

In our fiducial model presented in the previous section, star-forming galaxies at every halo mass were in equilibrium at $z=0$. As we mentioned, this is not necessarily the case in the real universe - the mass loading factor may well scale weakly with halo mass, in which case low-mass galaxies, with their long depletion times, would be unable to equilibrate to their baryonic accretion rate even in a Hubble time. Either scenario is currently perfectly consistent with observations, since the mass loss timescale is unknown, owing to its dependence on $\eta$. In figure \ref{fig:valid} we show regions of $M_h$-z space where the equilibrium assumptions are valid -- the bluer the color, the better-satisfied the condition that the Hubble time be much longer than $t_{eq}$, the maximum of the mass-loss time ($t_\mathrm{loss}$), the coherence time ($t_\mathrm{coherence}$), and the burn-through time ($\sigma t_\mathrm{loss}$).

To construct these diagrams, we also need to make an assumption regarding how input parameters vary not only with halo mass, but with redshift. We assume that the mass loading factor and the factor by which we reduce the efficiency, $f_\epsilon$, are independent of redshift, but that the depletion time scales as $t_{dep} \propto (1+z)^{-1}$, a somewhat weaker scaling than if the depletion time scaled with the dynamical time \citep{dave2012}. This scaling is consistent with recent observational results from CO observations at high redshift \citep{tacconi2010,tacconi2013, saintonge2013}, though of course there are large uncertainties. Moreover, these observations span a very limited range of mass and redshift compared to that shown in these diagrams. We therefore emphasize that these plots represent plausible assumptions, not predictions.



\subsection{Evolution with redshift}

Since it is plausible that a statistical equilibrium model may be used successfully at higher redshift, one may consider extending our analysis beyond the $z\approx 0.1$ data we have considered. There are however both theoretical and observational challenges. First, a great deal of poorly-justified assumptions are required to scale the fiducial model to higher redshift. Observationally, while there are measurements of the MS, MZR, and FMR at higher redshift, the metallicity measurements in particular are fraught with complications arising from the changing set of lines visible from ground-based telescopes and the uncertainty of converting line characteristics to physical metallicities in the substantially different environments of high-redshift galaxies. The high-order quantities we discuss in this paper are therefore both difficult to predict and measure.

We can nonetheless make the basic point under the assumption that the star formation rate and metallicity have reached their equilibrium values, the metallicity will be $Z_{eq} = Z_{IGM} + q$, where $q$ is a combination of $f_R$, $\eta$, and $\xi$, none of which are expected to change dramatically with redshift. Thus the MZR should not change with redshift if galaxies are in equilibrium. Observational studies tend not to find this result. However given the difficulty in calibrating the zero-point of the MZR even at low redshift, and the possibility that samples of high-redshift galaxies are biased high in SFR and therefore low in Z \citep{stott2013}, we do not believe equilibrium models have been ruled out at higher redshifts.

 Similarly, the slope of the FMR, which one may read off from the quadratic fit to the FMR (equation \ref{eq:quadratic}), 
\begin{equation}
\partial \log Z/\partial \log \dot{M}_\mathrm{SF} = a_2 + a_4 m + 2 a_5 s
\end{equation}
should be independent of the star formation rate. We therefore predict that in general $a_5\approx 0$, although since $m$ and $s$ are highly correlated, the fact that many best fit models do not yield $a_5=0$ is not necessarily an indication that this prediction is wrong. 

We can also describe how the fiducial model, which does not quite fit the data at $z=0$, would scale to higher redshift. The predicted value of $\sigma$ is not dependent on redshift, at least for the dark matter. Given our results suggesting that baryonic processes likely play an important role in smoothing the accretion, it is unclear how this smoothing process would evolve, so this constant value of $\sigma$ is highly uncertain. Meanwhile $t_\mathrm{coherence}$ will evolve strongly with redshift, but the quantity which sets the second-order scatter we consider is $\tau_c = t_\mathrm{coherence}/t_\mathrm{loss}$. If we assume $\eta$ at fixed halo mass does not change, then $t_\mathrm{loss}$ simply scales as the depletion time, which likely does decrease significantly towards higher redshift. If they decrease at a similar rate, near $1/(1+z)$, then $\tau_c$ is unlikely to evolve very much. In this scenario, we would predict that the scatter in the MS and MZR, and the slope in the FMR at fixed stellar mass would be roughly independent of redshift (and as evidenced in figure \ref{fig:contours}, independent of halo mass).

\subsection{Relationship to other work}
Our model bears a resemblance to several recent papers on equilibrium models \citep{dave2012,lilly2013}. Our model reduces to a slightly simpler version of the \citet{dave2012} model when $\tau_c \rightarrow \infty$ and $\sigma\rightarrow 0$, i.e. the upper left of the $\sigma-\tau_c$ diagrams we showed in sections \ref{sec:ms} and \ref{sec:Z}.

One of the criticisms of the \citet{dave2012} model has been its explicit assumption that $d M_g/dt = 0$. Simple toy models of the growth of galaxies under various star formation laws, \citep[e.g.][]{feldmann2013} point out that for many galaxies at redshift zero, $d M_g/dt < 0$. Indeed in our recent work on the radially-resolved evolution of disk galaxies since $z=2$ \citep{forbes2013}, we find that much of the galactic disk for many galaxies tends to be moderately out of equilibrium between local sources and sinks, with star formation being somewhat higher than the (local) accretion rate.

The equilibrium model of \citet{lilly2013} attempts to address this issue by allowing part of the incoming accretion to build up in the gas reservoir. The price they pay is that the star formation rate becomes an input to their model rather than an output. Perhaps even more worrying is that they assume $d Z/dt =0$ always, even when $d M_g/dt \ne 0$. Note that \citet{feldmann2013} makes a similar argument. As shown by our sample trajectories in section \ref{sec:Z}, it actually takes $Z$ longer to equilibrate than the star formation rate, since the metallicity can only equilibrate once the star formation rate catches up (or falls off) to the accretion rate. It is therefore odd that they assume the metallicity is in equilibrium while the star formation rate is not. It is only through this oddity that they are able to fit a second order relation, i.e. the FMR, with their model. We consider our model to be both more self-consistent and more powerful, in that we can generate a scatter in the star formation rate and metallicity, and about the FMR itself.

Another important result, and to our knowledge the only previous theoretical attempt to address the scatter in the main sequence, is \citet{dutton2010}. They use a rather sophisticated semi-analytic model, including cooling from virial shock-heated gas in a dark matter halo and star formation as a function of specific angular momentum (i.e., radius) in the disk, although they do not include any way for the gas to change its angular momentum \citep[as we do in][]{forbes2012,forbes2013}. They find a significant but small scatter in their model star-forming main sequence arising from variation in halo concentration, which in turn causes differences in the mass accretion histories between different galaxies of the same halo mass. Their model therefore resembles ours in the limit that $\tau_c \rightarrow \infty$, but $\sigma \ne 0$. Our model's more flexible treatment of the accretion process and other model parameters (e.g. the mass loading factor) gives us somewhat more insight on the problem of scatter not only in the main sequence but also in the MZR and FMR, although of course our model is far simpler in terms of its treatment of star formation, and we can make no predictions regarding other important and interesting quantities \citep[][i.e. galaxy sizes and rotation curves]{dutton2012}.

\section{Conclusion}
\label{sec:conclusion}

In the past few years, a new view has emerged as a useful way of understanding galaxies. In this picture, galaxies are in a slowly evolving equilibrium between accretion, star formation, and galactic winds regulated by the mass of cold gas in their interstellar media. To the degree that the parameters controlling this balance are well-defined functions of the mass of a galaxy and its redshift, this sort of model may be used to understand the connection between galaxy scaling relations and these physical parameters, which are not known from first principles.

In the spirit of these equilibrium models, we have presented a simple model which relaxes a key assumption in the equilibrium model, namely that the rate at which baryons enter the gas reservoir varies slowly. A population of galaxies in our model has been fed by the same stochastic accretion process for eternity, or at least long enough that the full joint distribution of all galaxy properties has become time-invariant. We therefore refer to our picture as a statistical equilibrium model, since the individual galaxies are not in equilibrium, but the population is.

With this model, we study a number of second-order relationships about the well-known galaxy scaling relations between the stellar mass and the star formation rate (the star-forming main sequence), and the stellar mass and metallicity (the mass metallicity relation). We look at the scatter at fixed stellar mass in both of these quantities, as well as the (anti-) correlation between star formation and metallicity at fixed stellar mass. Our main conclusions are as follows.

\begin{enumerate}
\item Including a stochastic scatter in the accretion rate at the level expected from N-body cosmological simulations naturally produces a scatter in both the star forming main sequence and mass-metallicity relation somewhat larger than the observed scatters. The anti-correlation observed between star formation rates and metallicities at fixed stellar mass is also naturally reproduced.
\item Neglecting the scatter in model parameters, (i.e., the mass loading factor, the depletion time, the scatter in stellar mass at fixed halo mass, etc.) all second-order quantities (the scatter in the main sequence, the scatter in the mass metallicity relation, and the slope in metallicity with respect to the star formation rate at fixed stellar mass) are determined by only two parameters: the scatter in the accretion rate, and the ratio of the timescale on which the accretion varies to the timescale on which the galaxy loses gas mass.
\item Using a maximally conservative interpretation of the available data, we are able to constrain these two parameters as well as a number of ``nuisance'' parameters, namely the scatter in the mass loading factor at fixed halo mass and the uncorrelated scatter in $M_*$ at fixed halo mass. We find that the log-normal scatter in the baryonic accretion rate is between about 0.1 and 0.4 dex, moderately smaller than what we would have predicted based on N-body simulations and assuming the baryons follow the dark matter. This may point to some process in the halos of galaxies which smooths out variations in the baryonic accretion rate, or a substantial amount of baryon cycling, which has the effect of averaging out the accretion rate over a longer time period. We find that the scatter in the mass loading factor is less than $0.1$ dex, remarkably small considering the theoretical uncertainty in the details of the physics of feedback. Our constraint on the timescale over which the accretion rate varies is much weaker, but could be narrowed considerably by stronger constraints on the Fundamental Metallicity Relation.

\end{enumerate}

We hope that the framework we have presented here motivates new development in both theory and observation. On the theory side, $t_\mathrm{coherence}$ (or a more sophisticated quantity describing the timescales on which accretion varies) and $\sigma$ may be measured with some confidence in both dark matter-only and baryonic cosmological simulations. Semi-analytic models may be altered to include the appropriate level of variability in baryonic accretion rates. Meanwhile, we have shown that observable quantities, e.g., the scatter in the star-forming main sequence, can provide significant constraints on properties of the baryonic accretion process and the galaxy-to-galaxy variability of the mass loading factor. Our inferences are, however, limited by our limited certainty on the intrinsic scatter in the scaling relations we have considered and the true parameters of the FMR. Pinning down these quantities observationally at a variety of masses and redshifts may substantially improve our understanding of the details of baryonic accretion and feedback.

\section*{Acknowledgments}
JCF is supported by the National Science Foundation Graduate Research Fellowship under Grant Nos. DGE0809125 and DGE1339067. MRK acknowledges support from the NSF through CAREER grant AST-0955300, and by NASA through ATFP grant NNX13AB84G. AB acknowledges support from the Cluster of Excellence ``Origin and Structure of the Universe''. AD was supported by ISF grant 24/12, by GIF grant G-1052-104.7/2009, by a DIP grant, and by the I-CORE Program of the PBC and The ISF grant 1829/12. The monte carlo simulations detailed in the appendix were carried out on the UCSC supercomputer Hyades, which is supported by NSF grant AST-1229745. We would also like to thank Erica Nelson, Pieter van Dokkum, Peter Behroozi, and Eyal Neistein for helpful conversations.

\bibliography{zotlib}

\appendix
\section{Details of the Monte Carlo Simulations}
\label{app:mc}

Throughout this paper we have presented heatmaps of various quantities as a function of $\tau_c = t_\mathrm{coherence}/t_\mathrm{loss}$, and $\sigma_{10} = \sigma\log_{10}(e)$. Computing each of these quantities for the model is typically a non-trivial task which requires a monte carlo simulation, in which a large ensemble of galaxies is sampled at random times to sample the underlying true distribution of the quantity in question for galaxies in this model. Here we describe the details of these simulations.

For each pixel in these grids of $\tau_c$ vs $\sigma_{10}$, we sample an ensemble of 30,000 galaxies. Each galaxy is started at a time $\tau=0$ with initial values $\Psi = Z^\dagger = 1$, the equilibrium values for those quantities in the limit $\sigma \rightarrow 0$ and $\tau_c \rightarrow \infty$. The galaxies are then evolved for long enough that, for all practical purposes they forget their initial conditions (formally our model assumes that the galaxy population has been undergoing the same stochastic accretion process for eternity, but this is obviously impractical computationally). To determine `long enough', we use the analytic results derived in sections \ref{sec:ms} and \ref{sec:Z} which show that galaxies forget their initial conditions with an e-folding time of $t_\mathrm{loss}$. We also note that for our distribution to represent the true long-term steady-state distribution, as discussed in section \ref{sec:applicability}, galaxies with large scatters in their accretion rate need to experience enough draws from the accretion rate distribution that even the tail of the probability distribution of past events has no influence on the present distribution.

We therefore define a timescale $\tau_\mathrm{long} = 1 + \tau_c + \sigma$, i.e. a time guaranteed to be of order the longest timescale in the problem for any choice of $\tau_c$ and $\sigma$. We then calculate the number of draws from the accretion distribution necessary to simulate each galaxy out to $15 \tau_\mathrm{long}$, namely $k = 15 \tau_\mathrm{long} / \tau_c$. We then draw a pseudo-random number uniformly distributed between 0 and 1, and compute the galaxy's properties ($\Psi$, $Z^\dagger$) at the time $\tau_\mathrm{obs} = (k+r) \tau_c $, where r is the random number. This samples the full distribution of these quantities for the population of galaxies in steady state. In general the computational cost is just proportional to $k$, since to compute $\Psi$ and $Z^\dagger$, we must first draw $k$ random numbers and compute the sequence $\Psi_1,...,\Psi_k$ and $Z^\dagger_1,...,Z^\dagger_k$ before we can calculate those quantities at $\tau_\mathrm{obs}$. Thus the models with short coherence times, i.e. $\tau_c \ll 1$, are the most expensive.

\begin{figure}
\includegraphics[width=9 cm]{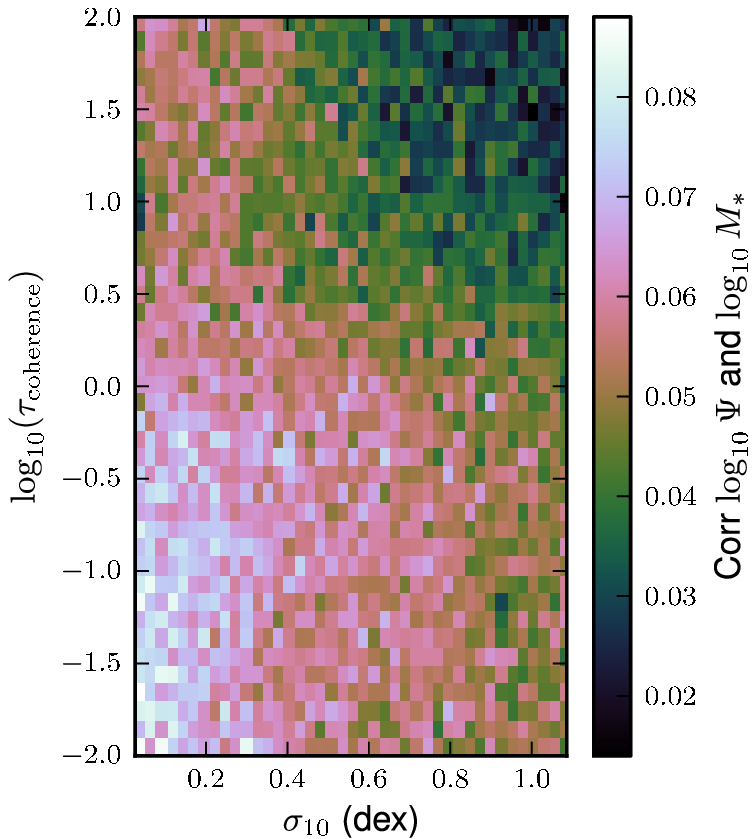}
\caption{The correlation between SGP versions of the star formation rate and stellar mass. The non-zero correlation shows that scatter in stellar mass at fixed halo mass can drive galaxies along the MS, rather than being merely uncorrelated.}
\label{fig:msf}
\end{figure}

\begin{figure}
\includegraphics[width=9 cm]{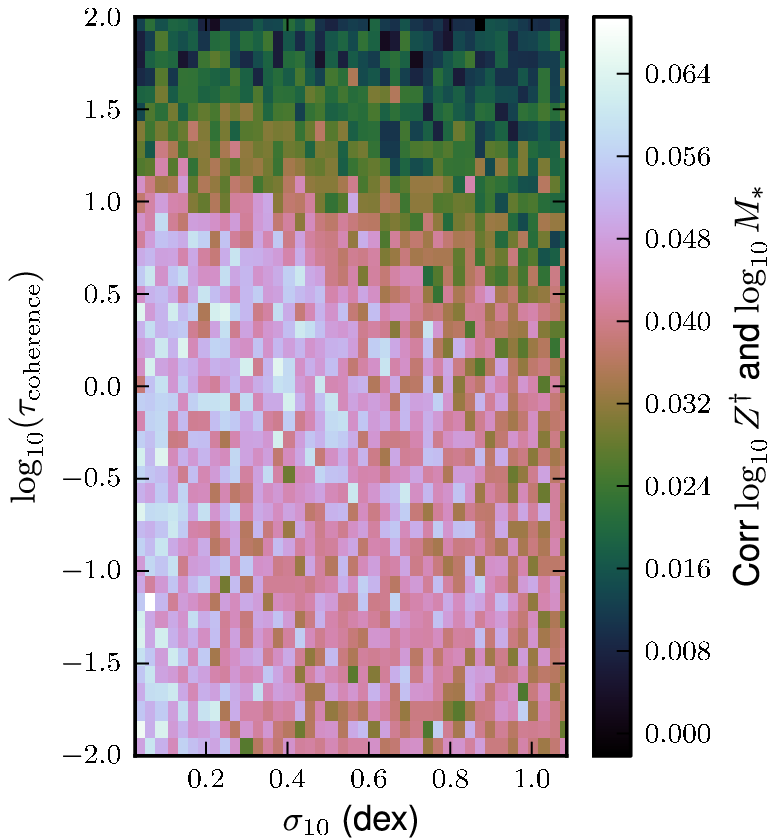}
\caption{The correlation between SGP versions of the metallicity and the stellar mass. Just as with the star formation rate, the correlation is positive everywhere, meaning again that scatter in stellar mass at fixed halo mass scatters galaxies along the MZR.}
\label{fig:mz}
\end{figure}

With the 30,000 samples for each of the 41 by 41 points in the grid of $\sigma$ and $\tau_c$, we can then compute each of the quantities shown in this paper -- the standard deviation of $Z^\dagger$ and $\Psi$ (independently), the correlation between the two quantities, and the linear slope. We also record a quantity which may be regarded as a proxy for stellar mass, defined as
\begin{equation}
\mathcal{M}_* = \int_0^{\tau_\mathrm{obs}} \Psi(\tau) d\tau
\end{equation} 
Naturally the magnitude of this quantity is, on average, proportional to the amount of time we let the simulations run, which is chosen subjectively to be $\ga 15 \tau_\mathrm{long}$. However, one may expect that any statistical properties which remove the mean may be physically relevant. In figures \ref{fig:msf} and \ref{fig:mz} we show the correlation between our analogs to star formation and metallicity, and stellar mass. Clearly, over much of parameter space there is a small but appreciable correlation between each quantity and $\mathcal{M}_*$, meaning that the tension between our result that $\sigma_{\log M_*} \la 0.15$ dex and the observational constraint that the scatter in stellar mass at fixed halo mass be 0.19 dex \citep{reddick2013}, is not a large concern. This is because our constraint is on scatter in $M_*$ that is uncorrelated with any other quantity, whereas in reality, as in the SGP, the stellar mass may well be positively correlated with the quantities in question, in which case at least some of the scatter in $M_*$ will be along the first-order scaling relations, and therefore won't contribute to the scatter in the relation at fixed stellar mass.

\clearpage

\end{document}